\documentclass[12pt,preprint]{aastex}
\usepackage{psfig}
\def\deg{$^{\circ}$}
\def\etal   {{\sl et~al.\ }}
\def\ph {$\phi$ Her}
\def\th {$\tau$ Her}

\begin{document}
\title{~~\\ ~~\\ The HgMn Binary Star $\phi$ Herculis: Detection and Properties of the Secondary and Revision of the Elemental Abundances of the Primary}
\shorttitle{$\phi$ Herculis}
\shortauthors{Zavala \etal}
\author{R.T. Zavala\altaffilmark{1}, S.J. Adelman\altaffilmark{2,3}, C.A. 
Hummel\altaffilmark{4}, A.F. Gulliver\altaffilmark{3,5}, H. 
Caliskan\altaffilmark{6}, \\ 
J.T. Armstrong\altaffilmark{7}, D.J. Hutter\altaffilmark{1}, K.J. 
Johnston\altaffilmark{8}, and T.A. Pauls\altaffilmark{7}}

\email{bzavala@nofs.navy.mil,adelmans@citadel.edu,chummel@eso.org}
\altaffiltext{1}{United States Naval Observatory, 10391 W. Naval Observatory 
Rd., Flagstaff, AZ 86001}
\altaffiltext{2}{Department of Physics, The Citadel, 171 Moultrie Street, 
Charleston, SC 29409}
\altaffiltext{3}{Guest Investigator, Dominion Astrophysical Observatory,
  Herzberg Institute of Astrophysics, National Research Council of
           Canada, 5071 W. Saanich Road, Victoria V8X 4M6 Canada}
\altaffiltext{4}{European Southern Observatory, Casilla 19001, 
                 Santiago 19, Chile}
\altaffiltext{5}{Department of Physics, Brandon University, Brandon, MB R7A 6A9
Canada}
\altaffiltext{6}{{\.I}stanbul University, Department of Astronomy and Space 
Sciences, 34452 University, {\.I}stanbul, Turkey}
\altaffiltext{7}{Remote Sensing Division, Code 7210, Naval Research Laboratory,
4555 Overlook Avenue, Washington, DC 20375} 
\altaffiltext{8}{United States Naval Observatory, 3450 Massachusetts Avenue, NW, 
Washington, DC 20392-5420}


\begin{abstract}

Observations of the Mercury-Manganese star \ph\ with the Navy Prototype Optical 
Interferometer (NPOI) conclusively reveal the previously unseen companion in 
this single-lined binary system. The NPOI data were used to predict a spectral 
type of A8V for the secondary star \ph\ B. This prediction was subsequently confirmed 
by spectroscopic observations obtained at the Dominion Astrophysical Observatory. 
\ph\ B is rotating at  50 $\pm$ 3 km s$^{-1}$, in contrast to the 8 km s$^{-1}$ lines of 
\ph\ A.  Recognizing the lines from the secondary permits one to separate them from 
those of the primary. The abundance analysis of \ph\ A shows an abundance pattern similar 
to those of other HgMn stars with Al being very underabundant and Sc, Cr, Mn, Zn, Ga, Sr, 
Y, Zr, Ba, Ce, and Hg being very overabundant.

\end{abstract}


\section{Introduction}

The Mercury-Manganese (HgMn) stars are a class of non-magnetic peculiar main 
sequence B type stars with effective temperatures between 10500 K and 15000 K.  
Their members show a wide variety of abundance anomalies with both depletions 
\citep[e.g., N,][]{roby} and enhancements (Hg: Leckrone et al. 1991; 
Mn: Woolf \& Woolf 1976) and tend to be slow rotators relative to their normal 
analogs. Enhancements of some elements may be as large as 10$^5$ times the solar 
abundances \citep{pres}. References to many recent abundance studies of HgMn stars can be
found by consulting, e.g., Adelman et al. (2004b, 2006). These abundance anomalies are 
thought to be produced in an extremely hydrodynamically stable environment from the 
separation of elements by radiatively-driven diffusion and gravitational settling \citep{mich}. 
Woolf \& Lambert (1999a) discovered three young HgMn stars in the Orion OB1 association.  
Two of these stars are members of the 1.7 Myr old OB1b sub-association which indicates 
how soon after the ZAMS such stars can be detected.  The HgMn stars are important 
laboratories for studying stellar hydrodynamical effects.  With results from many stars 
one can study the dependence of the elemental abundances on stellar parameters and 
make comparisons with theoretical predictions \citep{a01}.

$\phi$ Herculis (HD 145389, HR 6023, V=4.23, $\alpha_{2000}= 16^{\rm h} 08^{\rm m} 
46^{\rm s}$, $\delta_{2000}= 44\degr\ 56'$ 06'') is a HgMn star whose peculiar 
spectrum was noted in the Henry Draper catalog  \citep{hd} with an inverse 
dispersion of 159 \AA~mm$^{-1}$.  As most HgMn stars can only be found on
higher resolution spectra, this indicates that $\phi$ Her is a star strongly showing
the HgMn phenomena.  The Henry Draper catalog entry noted that the Ca II K line 
was nearly as strong as in $\alpha$ Canis Majoris. The excess in 
manganese was noted by \citet{m33}.  \citet{osawa} identified it as a HgMn 
star. It is also a single-lined spectroscopic binary \citep{aik,bab}, but the 
secondary has eluded detection until now. \citet{bm99} list \ph\ as a 
Hipparcos binary which was not resolved by speckle interferometry although an 
orbit based on the photocenter motion was computed \citep{hip-bin}. Adelman et al. (2001) 
performed the most recent spectroscopic analysis of \ph\ A with S/N $\ge$ 200 Reticon and 
CCD spectrograms. Commencing in 1997 interferometric observations of \ph\ with the 
Navy Prototype Optical Interferometer \citep[NPOI,][]{npoi} were made as part 
of a program to observe binaries in the catalog of \citet{bat}. These were 
spectroscopic binaries selected by one of us (Hummel) whose orbits were 
expected to be resolved by the NPOI. Observations with the NPOI continued 
through 2005 June, and the secondary was clearly detected in the visibility 
data \citep{zav}. The interferometric detection of the secondary allowed us to 
predict the secondary's spectral type. This was an important clue to the 
detection of the secondary's lines.  They are consistent with such a spectral
type using observations obtained at the Dominion Astrophysical Observatory. 

In this paper we describe the interferometric and spectroscopic observations 
which for the first time reveal the true nature of the secondary star of \ph. 
The current study found some broad-lines from the secondary.  Now we must 
correct for the contributions of the secondary in determining the properties 
of the primary star. These results show how the combination 
of interferometric and spectroscopic data allows for a single-lined binary to 
become a double-lined system, and hence improve our knowledge of the 
system.

\section{Observations and Data Reduction}

Both optical interferometric and spectroscopic data contributed to this 
study. We first consider the reduction of the data separately, and in 
\S3 apply the two together to test the prediction of the secondary 
star's spectral type and impact on \ph\ as a chemically peculiar star.  

\subsection{NPOI Observations}

\ph\ was observed with the NPOI on 25 different epochs from 1997 April 19 to 
2005 June 16. \citet{npoi} describe the NPOI in detail. Here we present a brief description of 
the instrument. The NPOI functions as a two to six element optical 
interferometer. The individual elements are siderostats with a 35 cm 
unvignetted aperture. The siderostats feed the light into evacuated pipes 
which send the light to a beam combining optical table. The pipes cause a 
stop-down of the effective aperture to 12 cm. Varying path length delays are 
removed by the use of Fast Delay Line (FDL) carts which are also contained in 
evacuated pipes. The FDL's move continuously to adjust for the changing delay 
caused by a varying projected baseline length due to the Earth's rotation. 
The light from the individual siderostats is combined on an optical bench to 
produce interference fringes. To provide an analogy to radio interferometry, 
the FDL's and the beam combining table serve as the NPOI's correlator. After 
combination spectrometers disperse the light which is collected onto lenslet 
arrays. These arrays feed the light to banks of avalanche photo diodes (APD) 
via fiber optic cables.     

\begin{figure}
\plotone{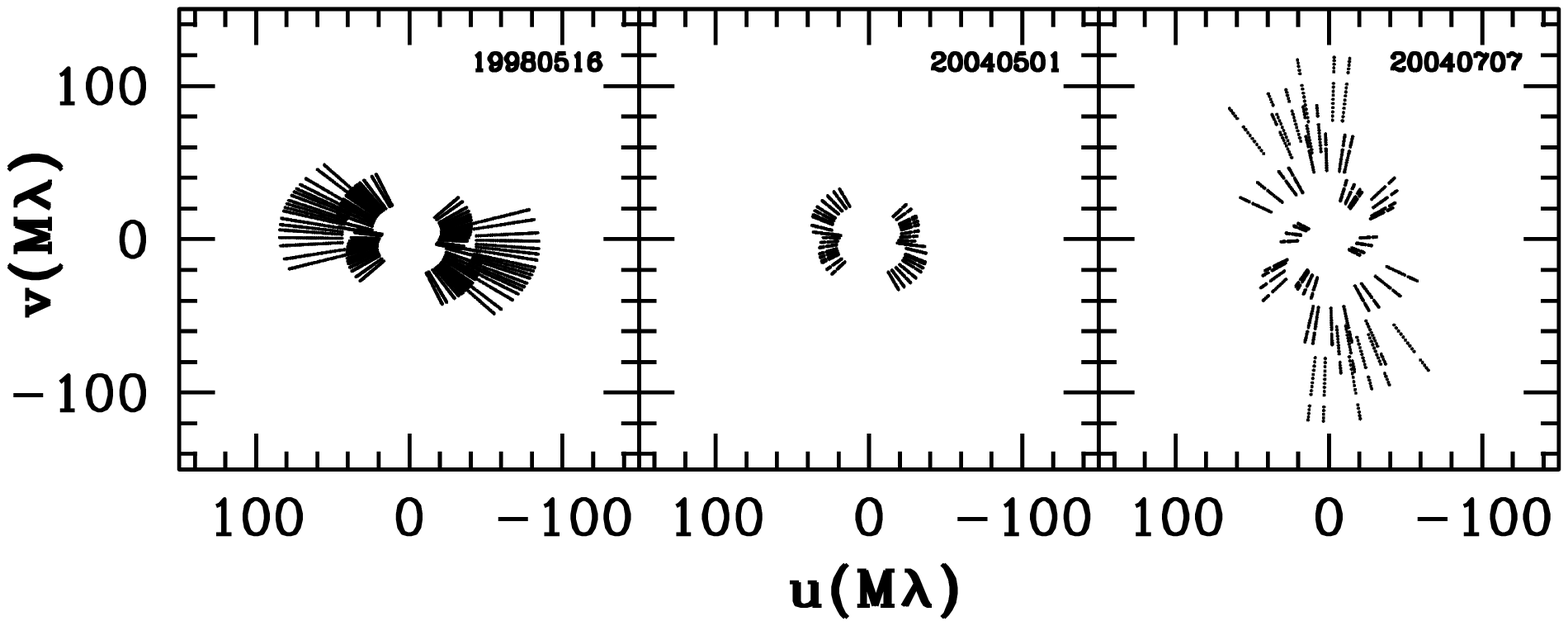}
\caption{Sample uv coverage plots for three epochs of NPOI observations 
of \ph. The epoch is indicated inside each subplot. Spatial 
frequency units in {\it{u}} and {\it{v}} are in mega wavelengths.}
\label{uv}
\end{figure}

NPOI observations before 2001 used 32 channels across a 4500-8500\AA\ 
bandwidth. The 2004 and 2005 observations spanned 5500-8500\AA\ and used 16 channels. 
Table~\ref{t1} provides a log of the NPOI observations. The change in number 
of spectral channels and wavelength coverage was required due to an 
improvement in the NPOI's beam combination instrumentation. These changes 
allowed for the detection of multiple baselines on a single spectrometer, 
increasing the number of available baselines and thereby improving the NPOI's 
ability to synthesize a filled aperture. The additional baselines more than 
offset the reduced 16 channel spectral resolution. See Fig.~\ref{uv} for an 
example of the increased sampling in the uv plane enabled by the 8 baselines of the 
2004 July 07 observation compared to the earlier observations. Several epochs in 
2004 did not take full advantage of this capability (e.g., those with 3 siderostats). 
These observations were conducted as part of a stellar multiplicity survey 
which necessitated the use of only two baselines \citep{tpf}. Details regarding 
the improvements which apply to the 2004 and 2005 observations can be found in 
\citet{etavir} and \citet{6spy}. For the 1999 and earlier observations 
see \citet{oldobs} for instrument details.

Observations of the target star \ph\ are interspersed with observations of the calibrator
star \th. \th\ is located 2.4\deg\ from \ph, has a V magnitude of
3.89 \citep{bsc}, and is a B5IV standard in the Revised MK system \citep{mkrev}.
 During an observation of either \ph\ or \th\ data are recorded every 2 msec. 
The 2 msec data are then averaged to produce points every 1 second. Data 
reduction was performed using C. A. Hummel's OYSTER software package. Data 
points were initially flagged on the basis of several factors. Outlier points 
in delay residuals, seeing indicators, photon rates and visibilities were 
removed. The 1 second data points were then averaged to produce squared 
visibility data for each 30 seconds (a ``scan'') taken of the calibrator 
and target star. These $V^2$ data contain an additive bias term and this bias was 
subtracted using the method described in \citet{etavir} and \citet{bias}. 

Calibration is performed using the expected angular diameter of the calibrator  
star \th. The color and apparent magnitude of a star can be used to estimate 
the uniform disk angular diameter \citep{diam,nat}. Using the R$-$I color of 
\th\ this diameter is estimated to be 0.28 mas. A multiplicative 
factor is then determined for each spectral channel so that the observed, 
bias corrected, $V^2$ data of the calibrator are brought inline with the 
theoretical expectation. This same calibration factor is then applied to the 
observed, bias corrected, $V^2$ data of \ph. Calibration factors were 
determined versus time, in the manner described for NPOI data in 
\citet{mizar}. 

\subsection{DAO Spectroscopy}

Our elemental analysis of $\phi$ Her A is an extension and modest revision of
that of Adelman et al. (2001) who obtained Reticon and CCD 
exposures with the long camera of Coud\'{e} spectrograph of the 1.22-m telescope
of the Dominion Astrophysical Observatory.  Seven new spectra were obtained with
the SITe4 CCD. The four centered at $\lambda$4864, $\lambda$5002, 
$\lambda$5140, and $\lambda$5278 are in the second order with a 
wavelength range of 147 \AA~and a two pixel resolution of 0.072 \AA.  Those 
centered at $\lambda$6562 and $\lambda$8556 are first order exposures with a 
wavelength range of 294 \AA~and a two pixel resolution of 0.144 \AA.   The
spectrum centered at $\lambda$4340 with a wavelength range of 394 \AA~and a two 
pixel resolution of 0.195 \AA~was a short camera exposure which was used to extract
the H$\gamma$ profile. The signal-to-noise ratios are respectively, 340, 300, 350, 
370,  330,  460, and 250  at the  continuum level.  No measurements were made in 
regions of very heavy telluric contamination.

The analysis techniques are very similar to those of Adelman et al. (2001). 
The stellar exposures were flat fielded with the exposures of an incandescent
lamp which was placed in the Coud\'{e} mirror train.  The separation  of the light of a 
desired order  involved both optical coatings of the Coud\'{e} mirror train and glass 
photographic filters.  To simulate the effects of the secondary
mirror, a central stop was used.  The one-dimensional scattered light corrected
spectra were extracted using the program CCDSPEC (Gulliver \& Hill 2002).

\section{Detection of the Secondary}

\subsection{Interferometry}

\begin{figure}
\plotone{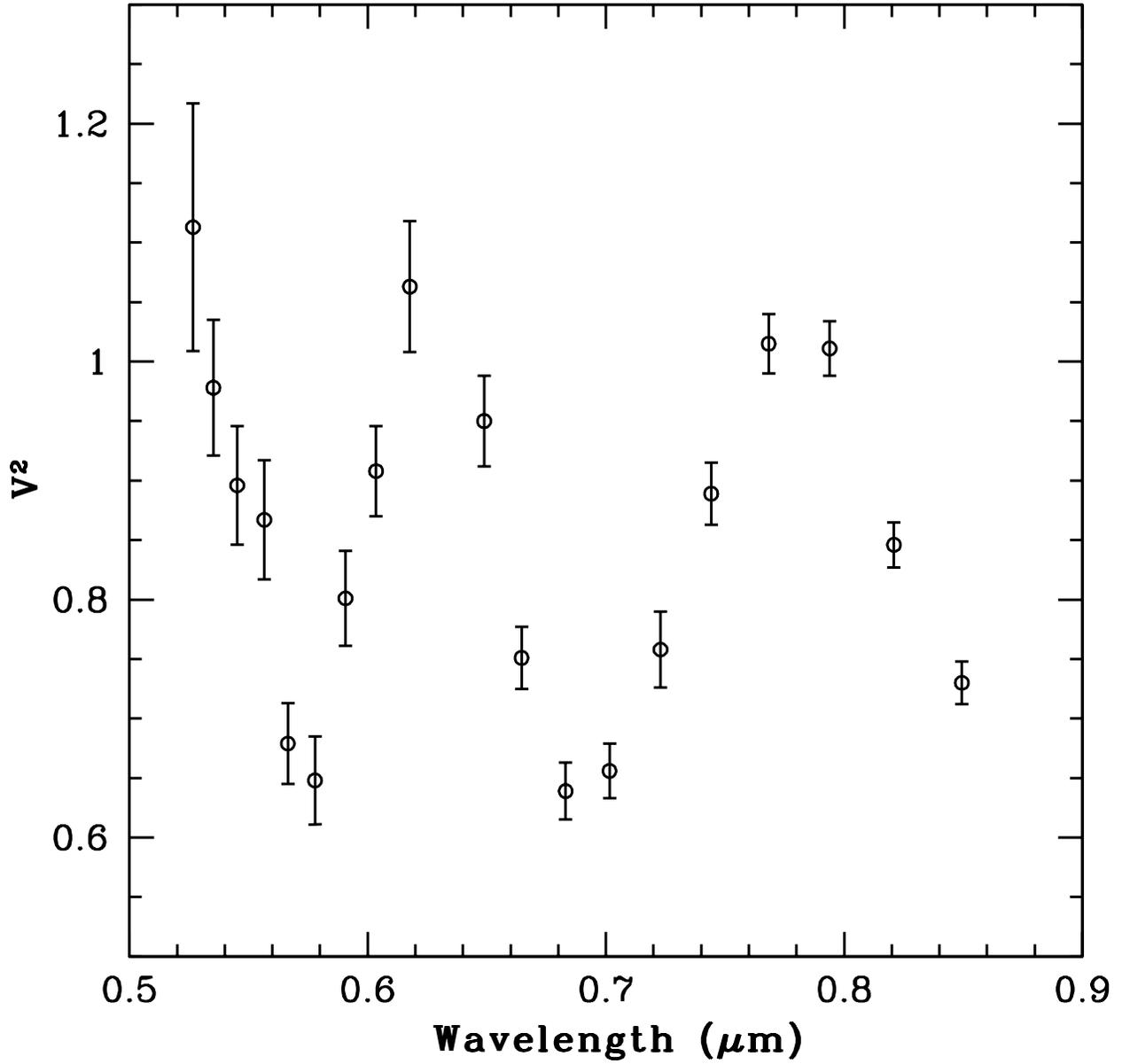}
\caption{Calibrated squared visibilities of \ph\ for scan 11 of the  
1998 May 16 NPOI observation. The data show the characteristic 
cosine wave signature of a binary. Data at less than 
0.526 $\mu$m were flagged due to poor sensitivity at the shorter 
wavelengths.}
\label{detect}
\end{figure}

As a single lined spectroscopic binary (SB1) one could assume that the secondary 
star was at least one magnitude fainter than \ph\ A \citep{heintz}. Other 
than that not much else was known of \ph\ B. There was a suggestion that the 
secondary could be a white dwarf \citep{wd}.
Fortunately optical interferometry allows the detection of secondary stars 
approximately three magnitudes fainter than the primary. \ph\ was observed with 
the NPOI as part of a program to detect the secondary stars in SB1 systems 
selected from \citet{bat}. Early observations with the NPOI easily revealed the 
characteristic oscillation \citep{pan} of a binary star in the squared visibilities 
of \ph. Fig.~\ref{detect} shows an example of the calibrated $V^2$ data 
versus wavelength for \ph\ obtained on 1998 May 16. 

With the calibrated $V^2$ data we can now estimate the separation $\rho$ and 
position angle $\theta$ of the binary for each night. The individual $\rho$ 
and $\theta$ values are then used to determine the seven orbital 
elements as for visual and speckle binaries \citep{heintz}. With the 
orbital elements we can use the visibility data from all observations and 
fit for the magnitude differences. Initial guesses for the stellar diameters and 
the magnitude differences are needed for the initial fits to $\rho$ and $\theta$ 
for individual nights. The primary's spectral type and parallax can be used to 
estimate a diameter for the primary. With a Hipparcos \citep{Hip} parallax of 
14.27 $\pm$ 0.52 mas a B8V star with a 3.0 R$_\odot$ radius \citep{allen} 
would have an angular diameter of 0.4 mas. The depth of the minimum $V^2$ is 
used to provide an estimate of the magnitude difference. Assuming both stars 
are on the main sequence we use the magnitude difference to estimate the secondary's 
spectral type and diameter using \citet{allen}. If both stars are on the 
main sequence we assume that \ph\ B's diameter is less than the diameter of A. 

\begin{figure}
\plotone{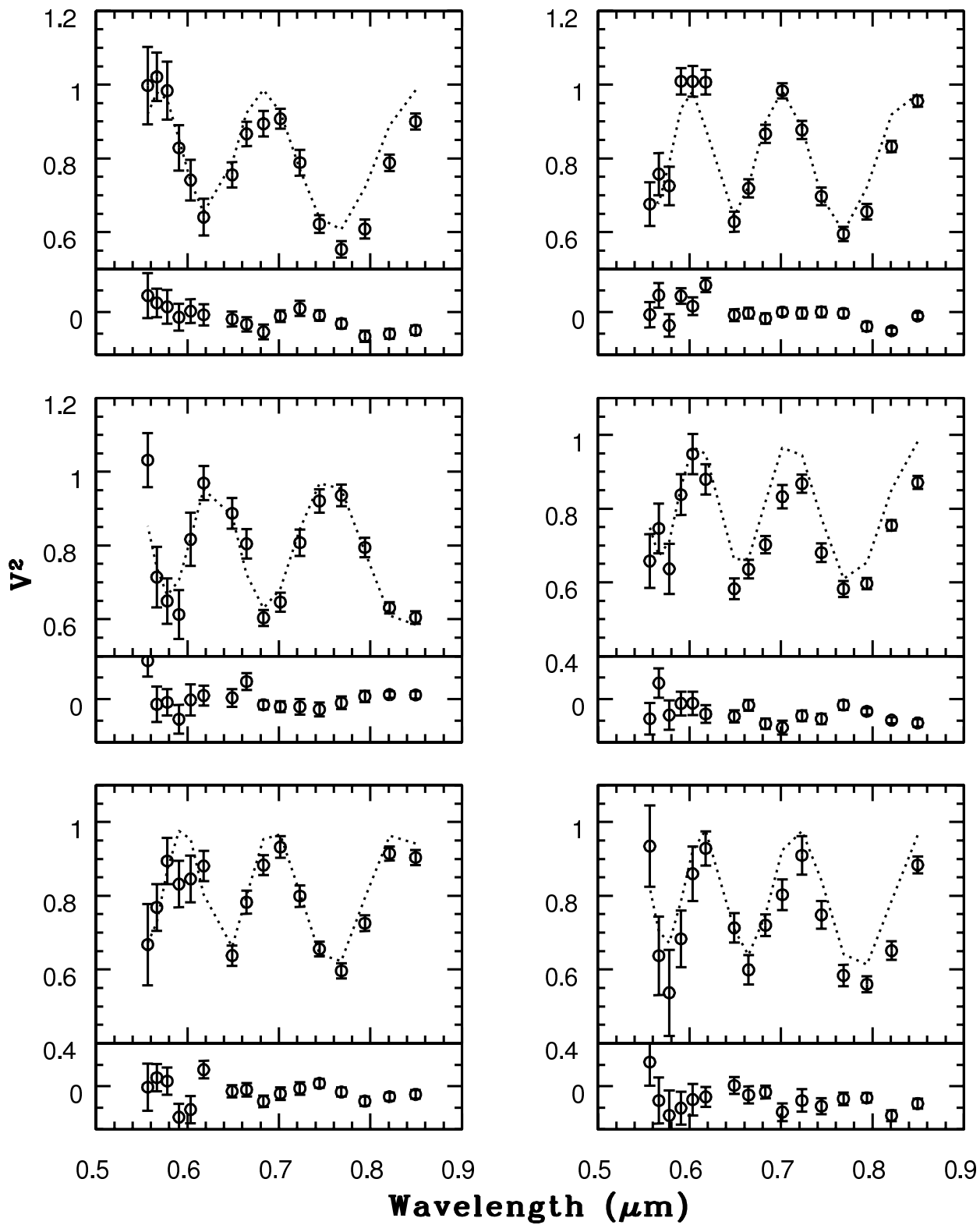}
\caption{Calibrated square visibilities of \ph\ are shown for 6 
scans for the 2004-07-31 NPOI observation. The model for \ph\ 
is shown as a dotted line. The residuals (calibrated $V^{2} -$
model) are shown for each scan in the small lower panels.}
\label{modfit}
\end{figure}

Fig.~\ref{modfit} shows an example of the final result for these fits. Six scans 
are shown for the 2004 July 31 NPOI data. The large panels show the calibrated
$V^2$ as open circles with the model shown as a dotted line. The small lower 
panels depict the residuals from the model fit. The residuals are larger at the 
blue end which is expected due to the NPOI's lower blue sensitivity relative to the 
red channels. In Table~\ref{astro} we show the relative astrometric results 
for \ph\ for all the NPOI observations. These relative astrometric solutions 
were used to solve for the orbital elements which are shown in Table~\ref{ephem}.

\begin{figure}
\plotone{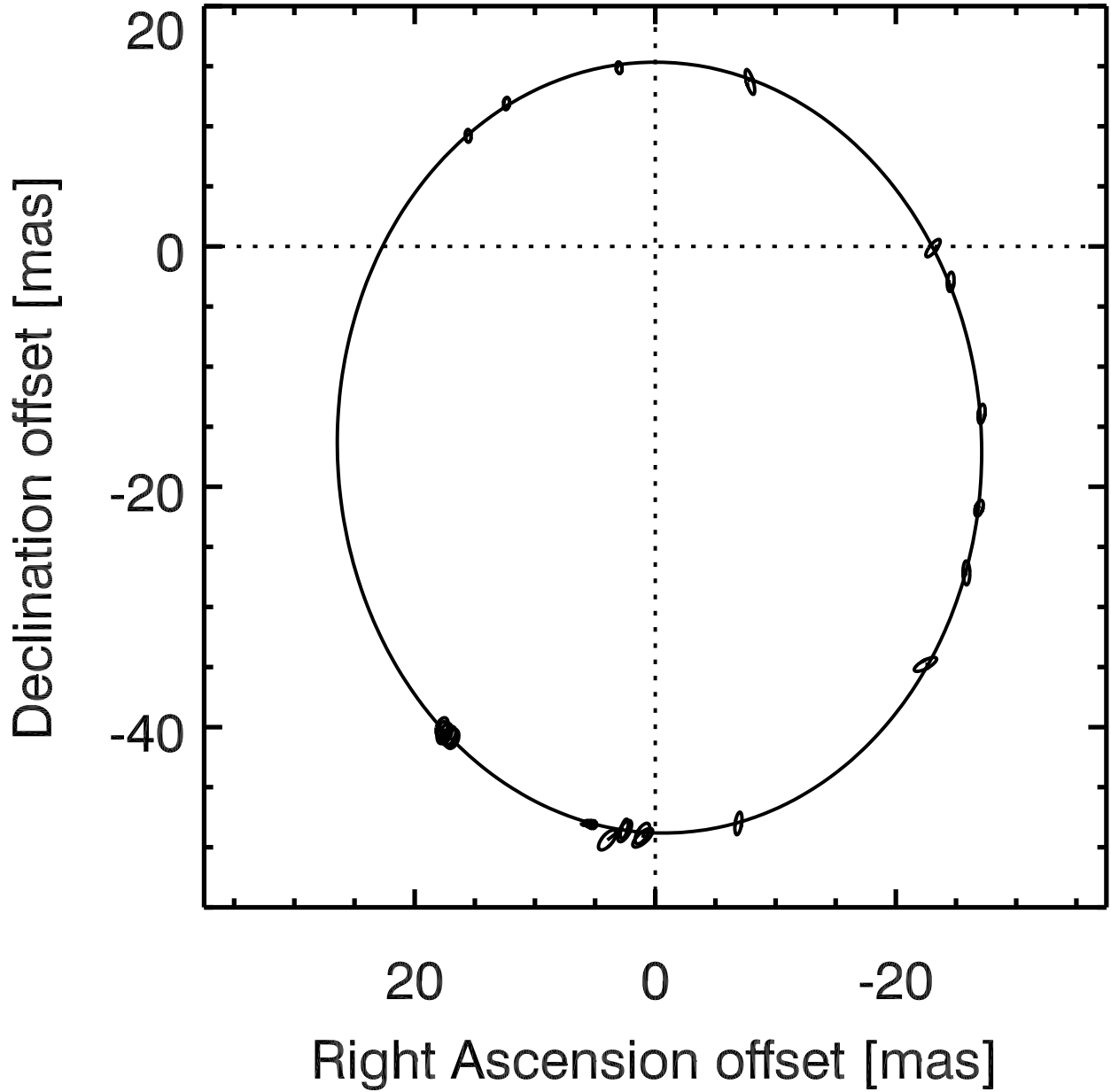}
\caption{The orbit of \ph\ using the elements of Table~\ref{ephem} plotted 
with the positions from Table~\ref{astro}. The primary is at the 
origin and the orbit represents the relative motion of the secondary
star. The dimensions and orientations of the error ellipses are given in 
Table~\ref{astro}. Vectors are occasionally visible which connect the observed 
position to the calculated position using the elements in Table~\ref{ephem}.}
\label{orbit}
\end{figure}

\begin{figure}
\epsscale{1.0}
\plottwo{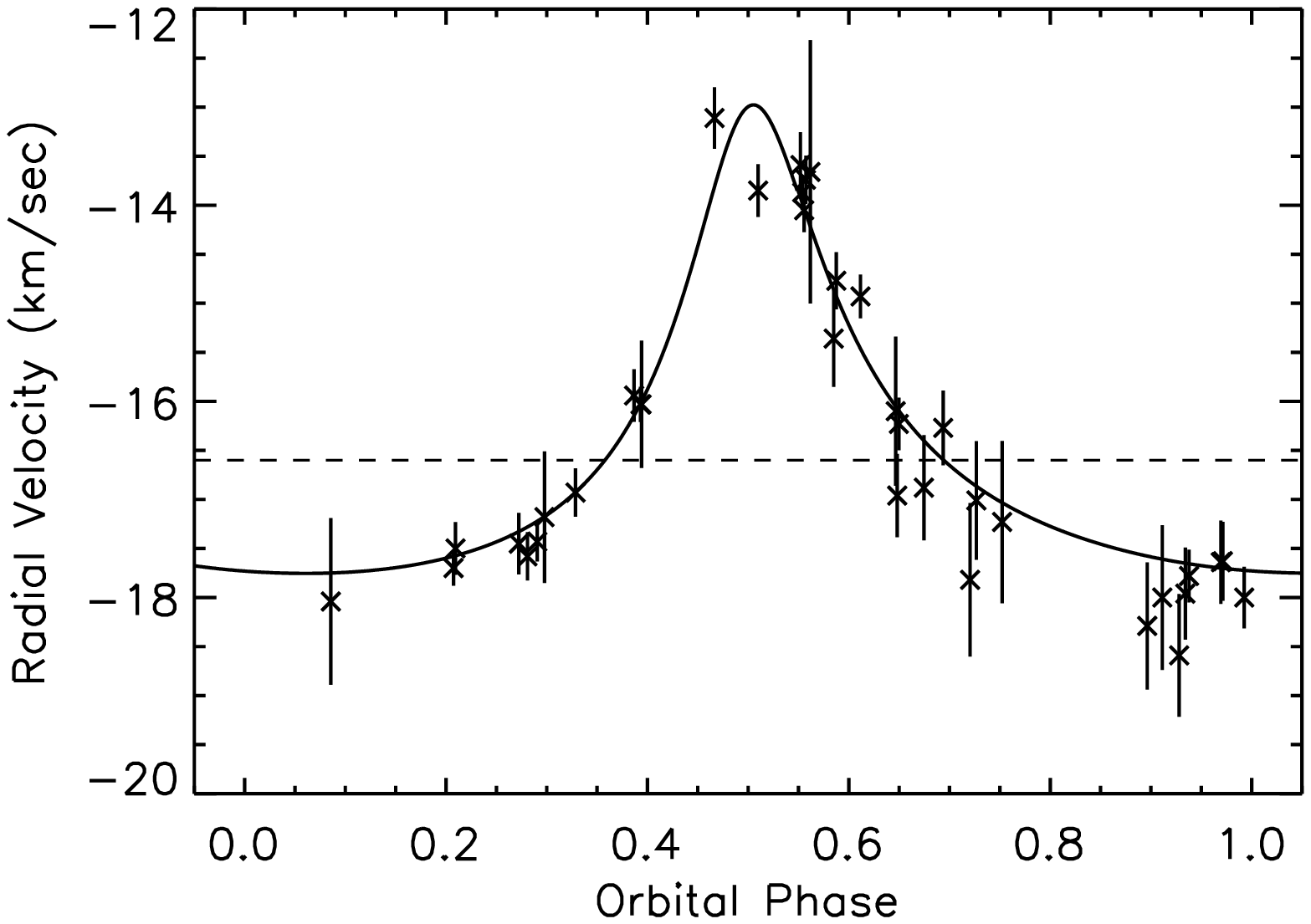}{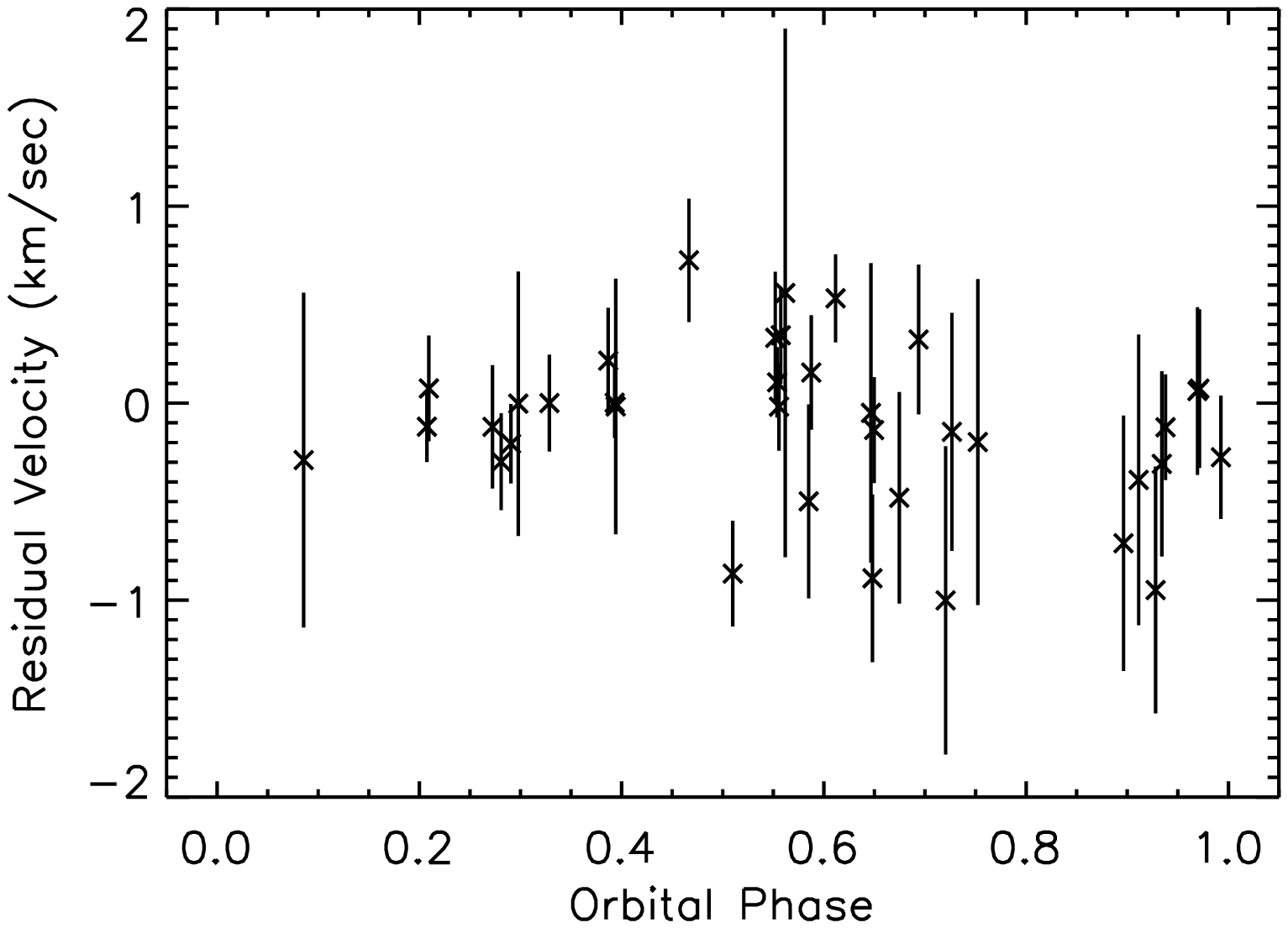}
\caption{Left panel shows the observed radial velocities of \ph\ A from \citet{aik} 
with errors increased by a factor of $\sqrt{5}$ plotted as a function of 
orbital phase. The solid curve represents the predicted radial velocities 
using the orbital elements in Table~\ref{ephem}. The right panel shows the 
residuals of the radial velocities again with errors of \citet{aik} 
increased by a factor of $\sqrt{5}$.}
\label{rvfig}
\end{figure}

\begin{figure}
\epsscale{1.05}
\plotone{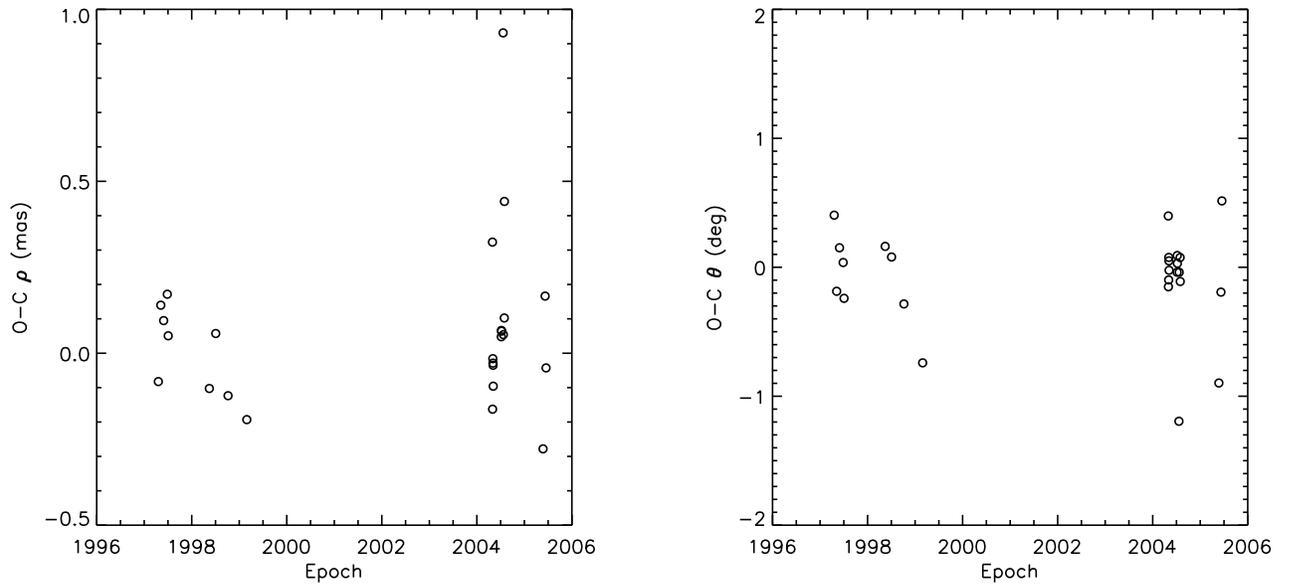}
\caption{Left panel shows the observed minus calculated  binary 
separation (O$-$C $\rho$) of \ph\ versus epoch of the NPOI observation. 
The right panel shows the observed minus calculated binary position angle 
(O$-$C $\theta$) of \ph\ versus epoch of the NPOI observation. The O$-$C 
data are listed in Table~\ref{astro} and were obtained using the 
orbital elements in Table~\ref{ephem}.} 
\label{oc}
\end{figure}

A combined solution of the orbital elements using the NPOI astrometric positions 
and the single lined radial velocity curve of \citet{aik} was performed.
An initial conservative estimate of the error of the astrometric results were 
the positions and orientations of the CLEAN beam, an elliptical Gaussian fitted 
to the FWHM of the dirty beam \citep{clean,cbb}. This combined radial velocity 
and astrometric solution resulted in a reduced chi squared of 3.4. The 
individual reduced chi-squared using only the astrometric data was 0.3, and using 
only the radial velocity data was 5.5. We elected to reduce the error ellipses of 
the astrometric results and increase the error estimates for the radial velocity data. 
The final error ellipses shown in Table~\ref{astro} have the dimensions 
of one-third of the CLEAN beam, and we increased the error estimates on 
the radial velocity data by a factor of $\sqrt{5}$. The subsequent 
orbital elements we fit have a reduced chi squared of 0.8 and are shown in 
Table~\ref{ephem}. The orbit overlaid on the astrometric results is shown 
in Fig.~\ref{orbit} and the radial velocity data and residuals are shown 
in Fig.~\ref{rvfig}. We also plot the observed minus calculated values 
for the separation and position angle contained in Table~\ref{astro}
and these are shown in Fig.~\ref{oc}. The orbital elements in 
Table~\ref{ephem} are significantly different from the Hipparcos 
derived orbit \citep{6orb,hip-bin}, and we did not incorporate the 
Hipparcos orbital elements in our solution.

\begin{figure}
\plotone{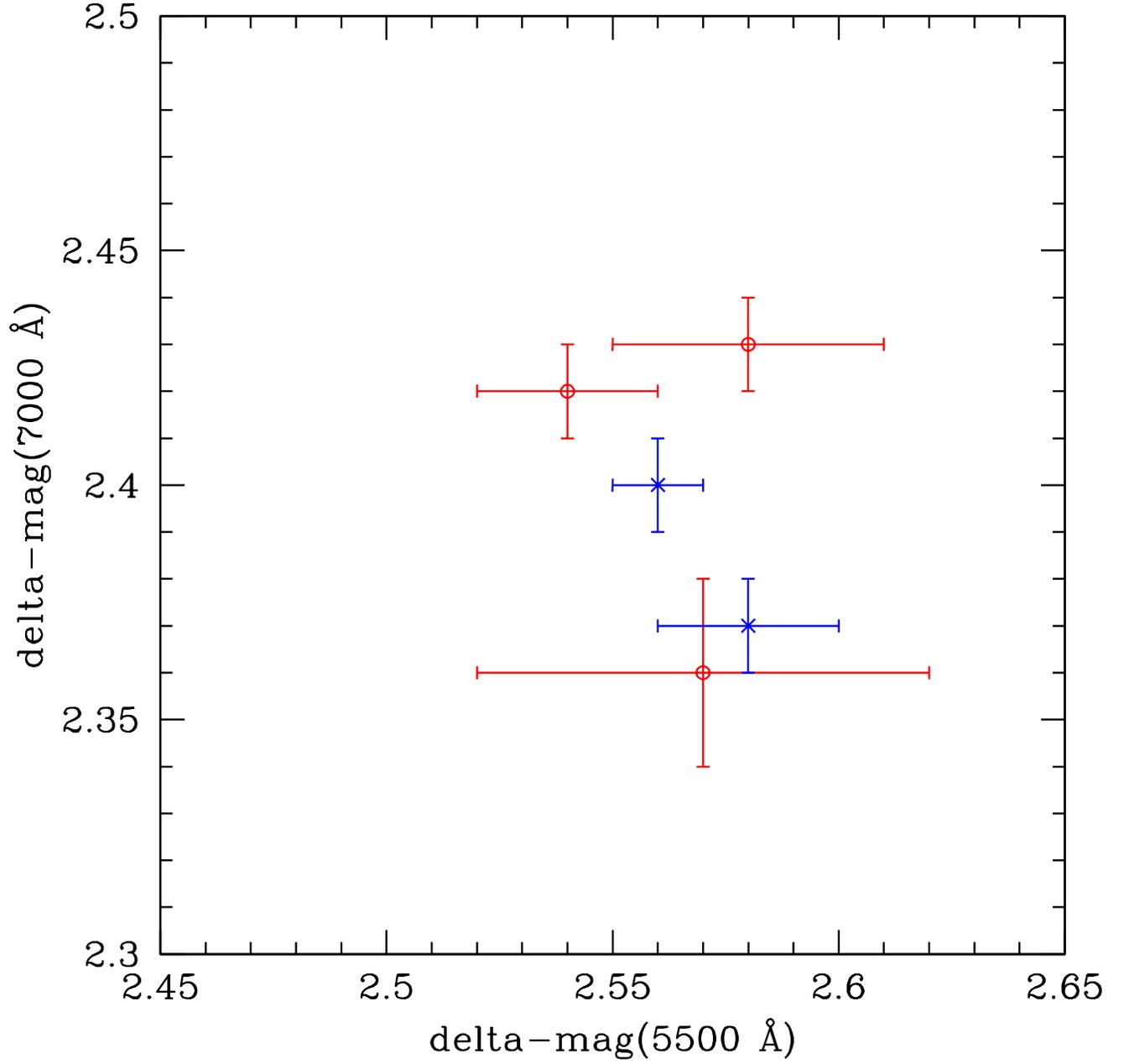}
\caption{The magnitude difference results for \ph\ from the NPOI data. 
The red circles show the 1997-1999 results, and the blue x's the 
2004 and 2005 results. Data for this figure are from Table~\ref{magdiff}.}
\label{delmag}
\end{figure}

The magnitude difference was the key to our prediction of the secondary 
spectral type, and we now consider the significance of that prediction.
The data were taken over several years with changes to the instrument 
configuration (see \S 2.1) so first we consider the magnitude difference 
results for the individual calendar years of our observations. Table~\ref{magdiff}
contains the results of fits to the magnitude difference at 5500\AA\ and 7000\AA\
for the 5 years for which the NPOI observed \ph. Fig.~\ref{delmag}
illustrates these results with the red points used for the 1997$-$1999 
observations and blue for the 2004$-$2005 observations. The scatter 
in Fig.~\ref{delmag} for the five different years of  
$\Delta$mag(7000\AA) is larger than our error bars. Also, if we ignore the one 
result for 1999 as an outlier the 2004 and 2005 observations suggest a 
smaller $\Delta$mag(7000\AA) compared to the 1997 and 1998 data. We decided to be 
conservative and use the maximum net extent of the 1$\sigma$ error bars in 
Fig.~\ref{delmag} to determine the magnitude differences and errors as: 
$\Delta$mag(5500\AA) = 2.57 $\pm$ 0.05 and $\Delta$mag(7000\AA) = 2.39 $\pm$ 0.05. 
The NPOI generally has a better sensitivity in the red as opposed to the blue 
end and we expect that systematic effects ultimately limit our errors to the level 
of $\pm$ 5\% for the magnitude difference independent of wavelength. 
Recent tests, conducted after 
these observations, show that our wavelength calibration for the 2005 observations 
may have been off by 2 to 8 nm in several channels. 
Residual calibration errors may also contribute to the error budget. Although 
these systematics are difficult to quantify 
Fig.~\ref{delmag} shows that across 7 years and with a change to the NPOI 
beam combination and fringe detection equipment the final error of $\pm$ 5\% 
in magnitudes seems rather good. With a magnitude difference of approximately 
2.6 in V an error of $\pm$ 0.05 magnitudes corresponds to an error in the 
measured squared visibilities of less than 3\%. 
 
The magnitude differences of 2.57 at 5500\AA\ and 2.39 at 7000\AA\
(V and R bands, respectively) allowed us to predict the secondary's spectral 
type as A8, using Table 15.7 of \citet{allen} assuming both stars are on the 
main sequence. Using the same table in \citet{allen} we see that the magnitude 
differences in V for a B8V$+$A5V binary is 2.2 and for a B8V$+$F0V binary it is 3.0. 
Both of these magnitude differences are at the 7$\sigma$ level for 
our NPOI results. This illustrates the NPOI's capability to determine 
the the spectral type of the secondary of \ph\ at the level of a few 
spectral subtypes.  

Using the Hipparcos parallax and the orbital elements in Table~\ref{ephem} 
we determined the sum of the masses to be 4.7 $\pm$ 0.6 M$_\odot$. 
The dominant source of error in the sum of the masses is the uncertainty 
of the parallax. If the parallax were perfectly known we would 
know the mass sum to approximately 3\%. 
The diameters used for \ph\ A and B in our model fits are listed in 
Table~\ref{ephem} but we did not attempt to fit a diameter for either star. 
The expected diameters for the two components from \citet{allen} are too small 
to resolve with the NPOI configurations used in this study.  

Using the mass sum from the dynamical parallax and the mass function of \citet{aik} 
of 5.4$\times$10$^{-4}$ we can estimate the individual stellar masses.  
This method gives 3.6 and 1.1 M$_{\odot}$ respectively for the primary 
and secondary. These masses are too large for a B8V and too small for an 
A8V \citep{and}. Using the tight mass$-$luminosity relationship and a 
bolometric correction of $-$0.75 for the primary we expect a B8V to 
have a mass closer to 3.1 M$_{\odot}$ which raises the secondary mass 
to 1.6 M$_{\odot}$. Clearly a precise determination of the stellar 
masses is not possible in this study of \ph. 
 
\begin{figure}
\epsscale{0.6}
\plotone{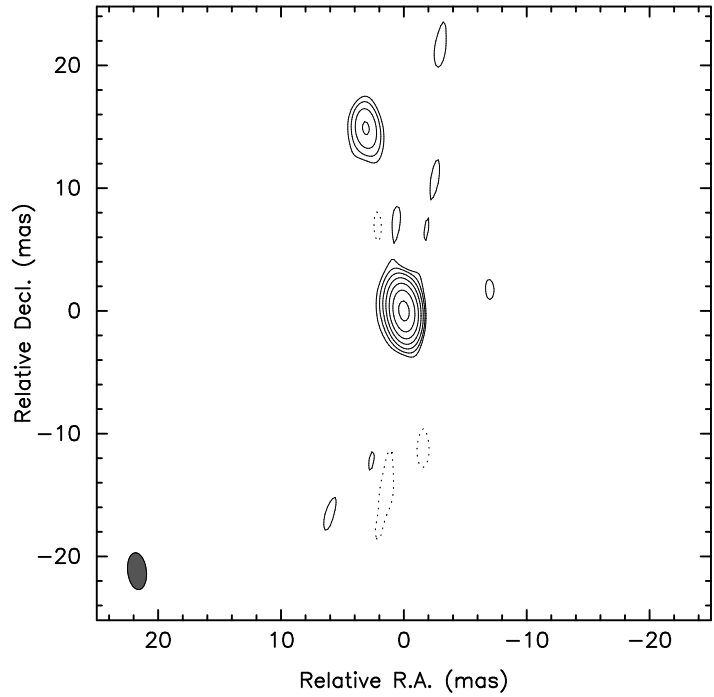}
\caption{Image of \ph\ on 2005 May 24 with a separation of 15 mas 
between components. Contour levels are -1.28, 1.28, 2.55, 5.1, 
10.2, 20.4, 40.8 and 81.6\% of map peak. The restoring beam 
is shown in the lower left hand corner and has dimensions of 
3.01 $\times$ 1.53 mas with a position angle of 6.52\deg.}
\label{image}
\end{figure}

An image of \ph\ produced from the 2005 May 24 NPOI observation is shown in 
Fig.~\ref{image}. The image was produced in DIFMAP \citep{shep,spt}
using the CLEAN algorithm \citep{clean} to remove the effects of irregular 
sampling in the uv plane (Fig.~\ref{uv}) to produce an image of the source. 
CLEAN requires as input the complex visibility \citep{tms} 

\begin{equation}
V \equiv |V|e^{i\phi}
\end{equation}

\noindent where $|V|$ is the visibility amplitude on a baseline and $\phi$
is the phase on that baseline. By taking the square root of the NPOI $V^2$
values and solving for the baseline phase with the requirement that the same 
closure phase is maintained we obtain the inputs required for CLEAN \citep{etavir}.
One limitation to this procedure is the requirement for at least three baselines 
which form a closure phase \citep{jenn} be present in the data. Thus the two 
baseline observations of \ph\ do not allow us to form an image of the binary.

\subsection{Spectroscopy}

The spectra were rectified using the interactive computer graphics program
REDUCE (Hill, Fisher \& Pockert 1982).  Gaussian profiles corresponding to 
v sin i = 8.0 km s$^{-1}$,  the value found by Adelman et al. (2001), were fit through 
most of the metal line profiles as the spectra were measured with VLINE which is 
part of the REDUCE package.   But a few lines were found 
corresponding to a rotational velocity of about 50 km s$^{-1}$. Rotational profiles 
were fit to these lines which must belong to the secondary, $\phi$ Her B. 
Thus $\phi$ Her now can be considered a  double-lined spectroscopic binary.

\begin{figure}
\epsscale{1.0}
\plotone{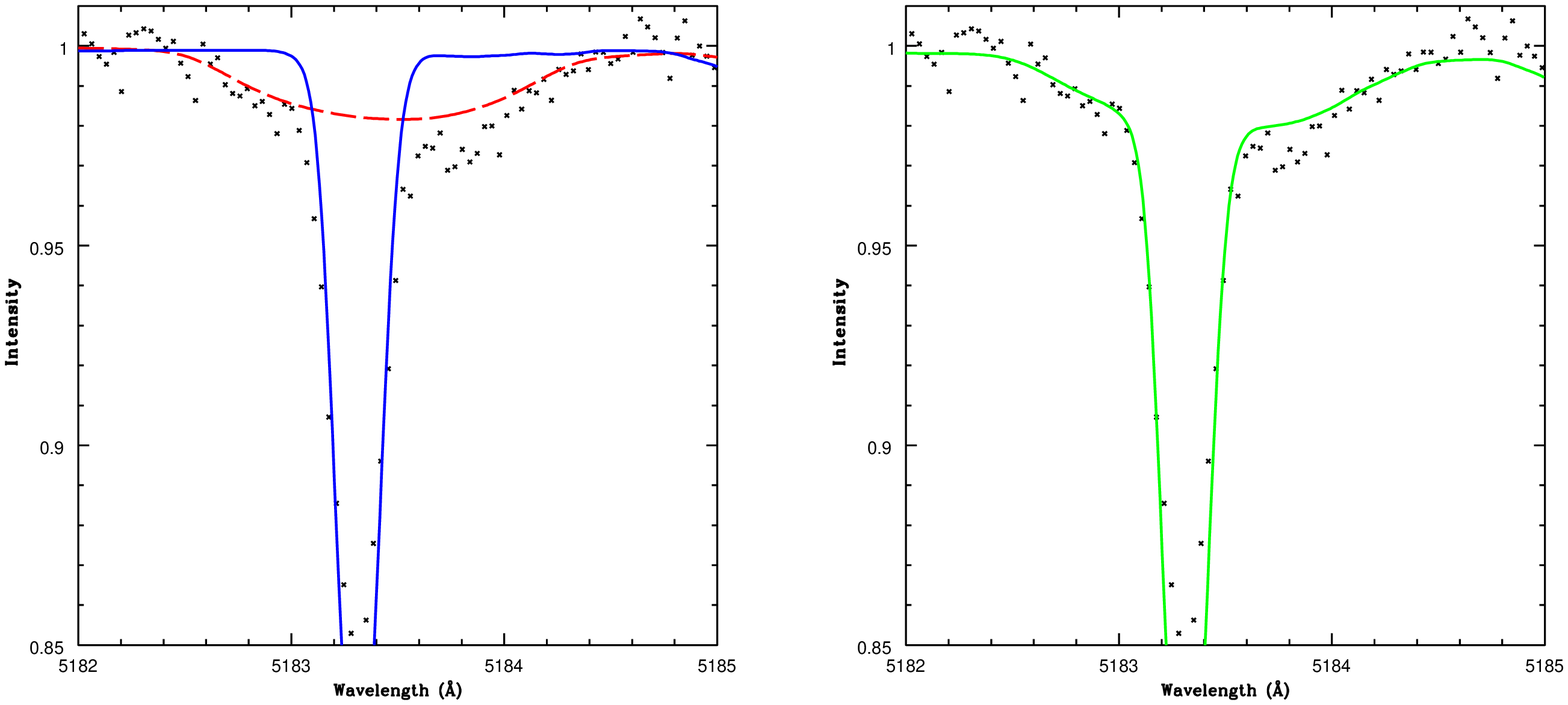}
\caption{Left panel shows our observed spectra (x) overlaid on normalized synthetic 
spectra for \ph\ A (blue solid line) and \ph\ B (red long dashed line) 
We can see the primary and secondary components of Mg I (2) $\lambda$ 5183.6042.
In the right panel we again show our data but now we overlay a line representing the 
summed synthetic spectra of \ph\ A and B. Further information on the epoch of observation 
is available in Table~\ref{rv-b}.}
\label{sum1}
\end{figure}

Measuring these features is not trivial.  It would be difficult to find them if
the S/N ratios were much less than 300 at the resolution of the DAO long Coud\'{e}
camera. Due to their shallow line depths, which are at most about 3\% below the 
continuum level, parts of their profiles can be poorly defined.  The resolution
 of our spectra is more than adequate to find them.  Some lines of $\phi$ Her B 
which are weaker than those detected might have been partially removed by the 
rectification process while others are lost in the continuum noise. A search of the
spectra of Adelman et al. (2001) did not reveal any additional candidates. Going to even 
higher S/N and to the red of $\lambda$5300 where the secondary 
contributes a greater percentage of light is the mostly likely way to find other lines in 
regions without telluric contamination. A list of those 
whose equivalent widths were greater than 10 m\AA~and whose profiles are sufficiently 
well defined for radial velocity measurements is given in Table~\ref{rv-b}.
Three of them are illustrated in Fig.~\ref{sum1}.  
The lines of $\phi$ Her B are about 2 \AA~wide while those due to the primary are much 
narrower. The synthetic spectrum in Fig.~\ref{sum1} was produced using a vsin(i) 
of 50 km sec$^{-1}$. We estimate the uncertainty in vsin(i) for \ph\ B 
as $\pm$ 3 km sec$^{-1}$ based on the 5 km sec$^{-1}$ error in the measured 
widths and the non-Gaussian line profiles.  

To identify the stellar lines we used the general references A Multiple Table of
Astrophysical Interests (Moore 1945) and Wavelengths and Transition
Probabilities for Atoms and Atomic Ions, Part 1 (Reader \& Corliss 1980) as
well as references for specific atomic species (see Adelman et al. 2001).  Zn II was 
the only new species we found.  The rest of the lines belong to those species found by
Adelman et al. (2001). For many species we found new lines to analyze.
We were only able to identify the strongest lines in the $\lambda$8556 spectrogram.

\begin{figure}
\epsscale{0.9}
\plotone{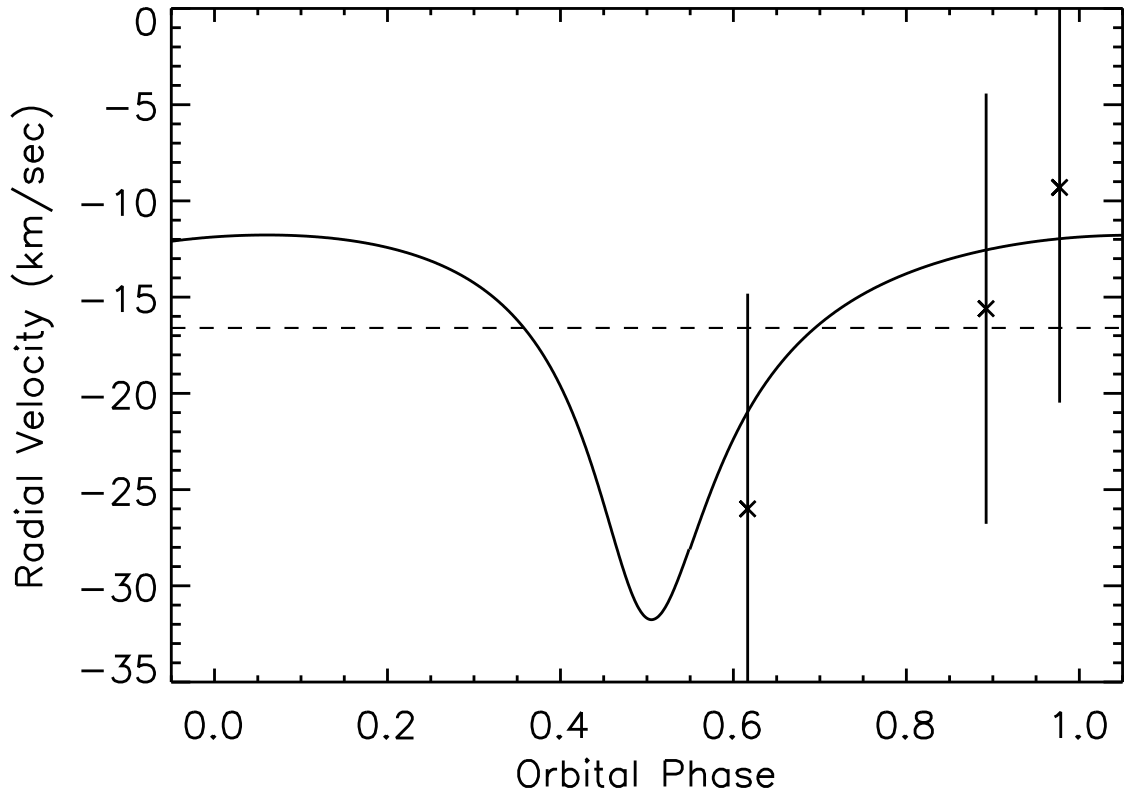}
\caption{The predicted radial velocity of the secondary of \ph\ B 
(solid line) using the orbital elements from Table~\ref{ephem} 
plotted with the observed radial velocities from Table~\ref{rv-b}.
The radial velocities from Table~\ref{rv-b} are averages of the 
velocities observed on 2004 Jun 10 and 2004 Jul 28.}
\label{rvb}
\end{figure}

We averaged the radial velocities for epochs 2004 Jun 10 and 2004 Jul 28, and plotted 
these averaged velocities with the epoch 2005 Jun 24 observation in Fig.~\ref{rvb}.
Using the averaged radial velocities from Table~\ref{rv-b} results in a negligible 
improvement to the reduced chi-squared. In \S3.1 we noted the difficulty in establishing a 
firm mass estimate. The uncertainties in the radial velocities of \ph\ B
do not provide a firm constraint on the velocity semi-amplitude. We note that 
our measured velocities are consistent with the prediction using the 
orbital elements in Table~\ref{ephem}. 

\label{s3}

\section{A New Look at \ph.}

To find the effective temperature and surface gravity of $\phi$ Her A, we used
as starting values those of Adelman et al. (2001).  Both their 20 \AA~mm$^{-1}$ spectrum
and our new 6.5 \AA~mm$^{-1}$ spectrum gave similar H$\gamma$ profiles.
Near the temperature of \ph\ A, the fluxes are almost independent of the surface 
gravity. To find the effective temperature of a single star one chooses a value of log g 
which is appropriate to its spectral type, assumes a value of the metallicity, 
and then calculates a grid of model atmospheres and predicted fluxes. By comparing the 
observations and the predictions, one finds the best fit which determines the 
effective temperature. After the abundance analysis is performed, this process may have
to be repeated so that the stellar and the model metallicities are similar. The 
effects of a non-solar metallicity are small. For a binary one can either add the 
predicted contributions properly weighted of the secondary (held constant) 
and of the primary (with the effective temperature varied) to predict the 
fluxes as observed or one can determine the energy distribution of the 
primary by subtracting the predicted contribution of the secondary from the 
observations. We chose the latter technique and used the spectrophotometry 
of the \ph\ system by \citet{ap83} and the fluxes predicted using the LTE 
plane-parallel ATLAS9 \citep{kur93} model atmospheres. The contribution of the 
secondary A8V star was assumed to be the fluxes from a T$\rm_{eff}$ = 
8000K, log g = 4.30 and solar metals model. \ph\ A has a T$_{\rm eff}$ = 11525K 
and log (g) = 4.05. 

Once the effective temperature is found for a single star, one compares a Balmer line 
profile (most often H$\gamma$) with the predictions of a series of models 
with the adopted T$\rm_{eff}$  and observed vsin(i) value over the most likely range 
of surface gravity. Then comparison of the observations and the predictions 
yields the surface gravity. For a binary star, one can either add the prediction 
for the secondary to that of the primary for comparison with observations 
or subtract from the observations the the prediction of the secondary to get the 
profile of the primary star. We used the former technique. If one has knowledge of 
the secondary star's radial velocity, this can be used in calculating the 
predictions.  The line profiles were calculated from the 
model atmospheres using SYNTHE \citep{kur81}.

The light contribution of the secondary to the joint spectra varies with wavelength, 
being for example, 7.9\% at $\lambda$4032 and 9\% at $\lambda$5360. One can 
correct the spectrum for its effects either before or after the measurement.
With rectified spectra, the correction amounts to removing the secondary spectrum 
assumed to be smooth from the joint spectrum and then renormalizing 
the remaining primary spectrum. For example, with the secondary spectrum 
amounting to 8\% the primary spectrum amounts to 92\%. So the correction scale 
factor is 1/0.92 or 1.087 and the line depths and equivalent widths increase 
by a factor of 1.087.  

\citet{lem} estimated uncertainties of $\pm$ 200K and $\pm$ 0.2 dex, when one uses 
calibrations of uvby-$\beta$ photometry to derive effective temperatures an surface 
gravities, respectively (see also Smalley \& Dworetsky 1995 for the results 
using fundamental stars). For results based on comparing optical region 
spectrophotometry and H$\gamma$ profiles with the predictions of models the 
uncertainties are slightly less. However, one can see differences in the fit 
which are much smaller than Lemke's error values. This strongly suggests that 
systematic errors may dominate these estimates. Errors in the measured absolute 
calibration of Vega in the optical region are 1\% at best \citep{hl75} and usually 
worse. For stars with parameters close to those of \ph\ A, the continuous energy 
distribution can be used for finding the temperature and then a Balmer line(s) for 
the surface gravity, any error in the former causes an error 
in the later. It is the second author's experience that when proceeding as we 
have in this paper that the errors in the absolute effective temperature and surface 
gravity are about 75\% of those quoted by \citet{lem} with the relative errors being much less.

We derived the helium and metal abundances using programs SYNSPEC
\citep{synspec} and WIDTH9 (Kurucz 1993), respectively, with metal line
damping constants from Kurucz \& Bell (1995) or semi-classical approximations
in their absence.  Abundances from Fe II lines were derived for a range of
possible microturbulences whose adopted values (Table~\ref{micro}) result in the 
derived abundances being independent of the equivalent widths ($\xi_{1}$) and 
having a minimal scatter about the mean ($\xi_{2}$) \citep{bss}.  We corrected 
the equivalent widths of metal lines of the primary for the contribution of the 
secondary. The errors in the equivalent widths as found by repeating the measurements 
many times is about 0.3 m\AA. We did not use Fe I lines as they were all on the 
linear part of the curve-of-growth and should have similarly shaped lines.  
Fe II has some lines which are stronger.  Our value for the 
microturbulence $\xi$ is typical of the HgMn stars.

The effects of errors in effective temperature and surface gravity on the metal
abundances are shown by Adelman et al. (2001) in their Table 3.
They found the changes in abundances due to a 100 K change in effective
temperature and a 0.2 dex change in log g. The sensitivities to effective
temperature are such that when the temperature is increased so are these
abundances, but for surface gravity often the neutral and singly-ionized
species have opposite dependencies.

The helium abundances were found by comparison of the line profiles with
theoretical predictions which were convolved with the rotational velocity and
the instrumental profile.  We corrected the line profiles for the contribution of the 
secondary which increases the He/H ratio slightly.  Adelman et al.  
(2001) found He/H = 0.06.  We could not derive the He/H ratios from any of 
the He I lines seen on the new  spectra due to blending. To convert 
log N/N$_{T}$ values to log N/H values -0.03 dex was added.

The analysis of the metal line spectra (Table~\ref{abun}) contains for each new line
the multiplet number (Moore 1945), the laboratory wavelength, the logarithm of
the gf-value and its source, the equivalent width in m\AA~as observed, and the
deduced abundance and the standard deviation about the mean.
For our abundance analysis, the errors of individual lines is of the 
order 0.20 dex (see, for example, Gigas 1986). Most astrophysical 
spectroscopists consider the total errors in the stellar mean abundances to be of order
0.30 dex. \citet{gig86} argues for errors of 0.22 dex in the best individual 
lines. We think our best mean abundances have errors slightly less than 0.30 dex.
The headers for each atomic species are given, but lines in Adelman et al. 
(2001) are omitted. For Mn II and Fe II, I and J, respectively, indicate the 
lines are from Iglesias \& Velasco (1964) and Johansson (1978).

Table~\ref{compare} compares the elemental abundances of \ph~A derived in 
this paper with those from Adelman et al. (2001) and from the Sun \citep{gns}. 
Most abundances changes between the two studies of \ph\ A are minor.  Our 
result for O I makes use of three multiplet 12 lines and only one of the two
previously analyzed lines. The scatter in the results from Mg I line results is much less 
due to the use of multiplet 2 lines rather than those from multiplet 3.  For Si II, 
S II, and Ni II to reduce the scatter  we excluded those lines most likely to be 
blended.  For Ba II, we used  $\lambda$4934 instead of $\lambda$4554 as 
it less likely to be blended and had a smaller abundance.

There are still some mild disagreements of results from elements with two or more atomic 
species, Mg, Ca, and Fe.  That the neutral species yield greater abundances 
than the first ionized species suggests that the stellar parameters still need 
a little more fine tuning.  One way to achieve this is to increase the stellar 
gravity a little. It is difficult to draw in a continuum over the H$\gamma$ 
line.  For the H$\gamma$ line profile to be within 0.5\% of the continuum, predictions based on 
ATLAS9 models \citep{kur79} show that for the effective temperature and 
surface gravity of \ph\ A one needs to look about 60 \AA\ from the line core to both 
shorter and longer wavelengths. Hence it is difficult to place 
the stellar continuum in the vicinity of this strong
Balmer line in A-type main sequence stars. There may also be minor problems with 
the derived energy distribution. But until the ASTRA spectrophotometer \citep{astra07} 
is operating, this possibility cannot be checked properly.  The lines from our new
spectra helped reduce the standard deviations of the mean for many species.

As \citet{wlam} use slightly greater resolution spectra than we have, do a proper 
analysis of the isotopic splitting for Hg II $\lambda$3984, and use similar 
values of T$_{\rm eff}$ and log g for $\phi$ Her A, we adopted their
result.  It also brings the Hg II result into agreement with that from Hg I $\lambda$4358.
For Nd and Pr we quote the values from Dolk et al. (2002) whose effective temperature
for $\phi$ Her A was 250 K greater than ours.  The lines used were often weak. We 
found some additional Nd III lines in its spectrum. 

\begin{figure}
\plotone{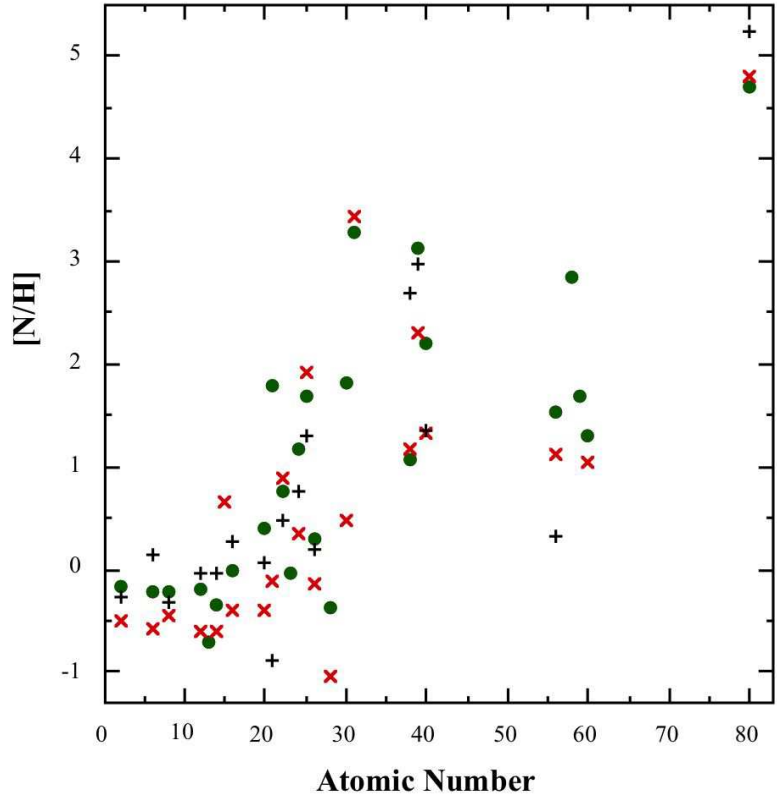}
\caption{The abundance anomalies [N/H] relative to solar for the HgMn stars
HR 7018 (+'s), $\upsilon$ Her (x's), and $\phi$ Her A (filled circles).  They show
the general pattern of anomalies for the HgMn stars along with individual scatter.}
\label{figanom}
\end{figure}

The abundance anomalies [N/H] = log N$_{H star}$ - log N$_{H sun}$ for $\phi$ Her A  
and for two sharp-lined HgMn stars $\upsilon$ Her (Adelman et al. 2006) and 
HR 7018 (Adelman et al. 2001) are shown in Fig.~\ref{figanom}   $\upsilon$ Her is a single star
while HR 7018 is a single-lined spectroscopic binary.   $\phi$ Her A has many 
of the characteristics of other HgMn stars.  The light elements in $\upsilon$ Her are 
slightly sub-solar and are either solar or  sub-solar in $\phi$ Her A. 
Scandium is very overabundant as are chromium, and manganese, while titanium
is overabundant, vanadium has a solar abundance, iron is slightly
overabundant, and nickel is underabundant.
Elements with atomic numbers greater than
28 are found to be overabundant when detected. 

\section{Conclusions}

Optical interferometric observations made with the NPOI conclusively detected 
the secondary star of the HgMn star \ph. The NPOI data enabled the prediction 
of a secondary spectral type of A8V. This prediction was confirmed via 
spectroscopic observations obtained at the Dominion Astrophysical Observatory. 
Until now lines of the secondary remained hidden in the generally much stronger 
lines of the primary. The secondary lines appear rotationally broadened to 
approximately 50 km s$^{-1}$, consistent with the rotation rates for normal 
A stars. This result does put to rest the tentative, though interesting, 
suggestion by \citet{wd} that the secondary  of \ph\ could be a white dwarf. 
Some changes to the abundances of \ph\ A are made and appear in 
Table~\ref{compare}. In Table~\ref{stars} we summarize 
the stellar parameters for the two components of \ph. This combination of optical 
interferometry and spectroscopy enables an increase in the number of chemically 
peculiar stars in double-lined binary systems. A detailed knowledge of the secondary 
star is required to completely remove its effect on the chemical abundances 
observed in the companion HgMn star. Lines most likely to be affected by 
blending with the secondary are identified and removed from consideration 
in the abundance analysis. 

\acknowledgments 

RTZ thanks Brenda Corbin, Gregory Shelton and Sally Bosken of the USNO Library for 
assistance with several references, and Brian Mason and Chris Tycner of USNO for
helpful discussions. This research has made use of the SIMBAD 
database, operated at CDS, Strasbourg, France, NASA's Astrophysics Data 
System Abstract Service, and the Washington Double Star Catalog maintained 
at the U.S. Naval Observatory. SJA and AFG thank Dr. James E. Hesser, Director of 
the Dominion Astrophysical Observatory for the observing time.  SJA's 
contribution to this paper was supported in part by grants from The Citadel 
Foundation.  Financial support was provided to AFG by the National Sciences 
and Engineering Research Council of Canada. HC thanks Dr. Dursun Kocer for his 
advice. The work done with the NPOI was performed through a collaboration 
between the Naval Research Lab and the US Naval Observatory, in association 
with Lowell Observatory, and was funded by the Office of Naval Research and 
the Oceanographer of the Navy. We thank the NPOI staff for the careful 
observations which contributed to this work, and our colleague Jim Benson 
for his efforts in support of the NPOI observations.   

{\it Facilities:} \facility{NPOI ()} \facility{DAO:1.2m ()}

\clearpage

\clearpage

\clearpage

\clearpage

\clearpage

\clearpage

\clearpage

\clearpage

\clearpage

\clearpage

\clearpage

\clearpage

\begin{deluxetable}{lccccr}
\tabletypesize{\scriptsize}
\tablecaption{NPOI O{\sc bservation} L{\sc og} \label{t1}}
\tablewidth{0pt}
\tablehead{
\colhead{UT Date} & \colhead{Julian Year} & \colhead{Sids} & 
\colhead{Baselines} & \colhead{MaxBl} & \colhead{Scans} \\
\colhead{(1)} & \colhead{(2)} & \colhead{(3)} & \colhead{(4)} 
& \colhead{(5)} & \colhead{(6)} }
\startdata
1997 Apr 19 & 1997.2972 & AC, AE, AW & 3 & 37.5 & 4 \\
1997 May 08 & 1997.3492 & AC, AE, AW & 3 & 37.5 & 3 \\
1997 May 30 & 1997.4094 & AC, AE, AW & 3 & 37.5 & 7 \\
1997 Jun 27 & 1997.4861 & AC, AE, AW & 3 & 37.5 & 5 \\
1997 Jul 04 & 1997.5052 & AC, AE, AW & 3 & 37.5 & 6 \\
1998 May 16 & 1998.3704 & AC, AE, AW & 3 & 37.5 & 18 \\
1998 Jul 04 & 1998.5046 & AC, AE, AW & 3 & 37.5 & 9 \\
1998 Oct 07 & 1998.7646 & AC, AE, AW & 3 & 37.5 & 4 \\
1999 Feb 28 & 1999.1589 & E2, E4, AW & 3 & 37.5 & 4 \\
2004 Apr 30 & 2004.3280 & AC, AE, AW & 2 & 22.2 & 2 \\
2004 May 01 & 2004.3307 & AC, AE, AW & 2 & 22.2 & 10 \\
2004 May 03 & 2004.3362 & AC, AE, AW & 2 & 22.2 & 16 \\
2004 May 04 & 2004.3389 & AC, AE, AW & 2 & 22.2 & 8 \\
2004 May 05 & 2004.3417 & AC, AE, AW & 2 & 22.2 & 7 \\
2004 May 06 & 2004.3444 & AC, AE, AW & 2 & 22.2 & 8 \\
2004 Jul 07 & 2004.5142 & AC, AE, AW, W7, AN & 8 & 66.5 & 8 \\
2004 Jul 08 & 2004.5168 & AC, AE, AW, W7, AN & 8 & 66.5 & 8 \\
2004 Jul 09 & 2004.5197 & AC, AE, AW, W7, AN & 8 & 66.5 & 8 \\
2004 Jul 21 & 2004.5525 & AC, AE, AW & 2 & 22.2 & 1 \\
2004 Jul 23 & 2004.5580 & AC, AE, AW & 2 & 22.2 & 13 \\
2004 Jul 30 & 2004.5771 & AC, AE, AW & 2 & 22.2 & 14 \\
2004 Jul 31 & 2004.5800 & AC, AE, AW & 2 & 22.2 & 7 \\
2005 May 24 & 2005.3929 & AC, AE, AW, AN, E06 & 6 & 53.2 & 3 \\
2005 Jun 09 & 2005.4368 & AC, AE, AW, AN, E06 & 6 & 53.2 & 4 \\
2005 Jun 16 & 2005.4561 & AC, AE, AW, AN, E06 & 6 & 53.2 & 1 \\
\enddata
\tablecomments{Col. (1): UT date of NPOI observation. Col. (2): Julian 
Year at 0700 UT (local midnight). Col. (3): Siderostats used. 
See \citet{npoi} for siderostat definitions and locations.
Col. (4): Number of baselines. Col. (5) Maximum baseline length (m). 
Col. (6): Number of interferometric scans obtained.}
 
\end{deluxetable}

\clearpage

\begin{deluxetable}{lccrccrrccc}
\tabletypesize{\scriptsize}
\tablecaption{A{\sc strometric} S{\sc olutions for} $\phi$ H{\sc er} \label{astro}}
\tablewidth{0pt}
\tablehead{\colhead{UT} & \colhead{Julian} & \colhead{$\rho$} & \colhead{$\theta$} & \colhead{$\sigma_{maj}$} 
& \colhead{$\sigma_{min}$} & \colhead{$\phi$} & \colhead{C$_\rho$} 
& \colhead{C$_\theta$} & \colhead{(O$-$C)$_{\rho}$} & \colhead{(O$-$C)$_{\theta}$} \\
\colhead{Date} & \colhead{Year} & \colhead{(mas)} & \colhead{(deg)} & \colhead{(mas)} & \colhead{(mas)} & 
\colhead{(deg)} & \colhead{(mas)} & \colhead{(deg)}& \colhead{(mas)} & \colhead{(deg)}  \\
\colhead{(1)} & \colhead{(2)} & \colhead{(3)} & \colhead{(4)} & \colhead{(5)} 
& \colhead{(6)} & \colhead{(7)} & \colhead{(8)} & \colhead{(9)} & \colhead{(10)} &\colhead{11}}
\startdata
Apr 19 & 1997.2972 & 37.51 & 223.59 & 1.00 & 0.28 & 181.0 & 37.59 & 223.19 & $-$0.08 &   \phs0.40 \\
May 08 & 1997.3492 & 34.63 & 231.01 & 0.69 & 0.31 & 163.0 & 34.49 & 231.20 & \phs0.14 &  $-$0.19 \\
May 30 & 1997.4094 & 30.50 & 242.80 & 0.81 & 0.28 & 172.0 & 30.41 & 242.65 & \phs0.09 &   \phs0.15 \\
Jun 27 & 1997.4861 & 24.73 & 263.16 & 0.81 & 0.28 & 175.6 & 24.56 & 263.12 & \phs0.17 &   \phs0.04 \\
Jul 04 & 1997.5052 & 23.09 & 269.66 & 0.89 & 0.30 & 141.7 & 23.04 & 269.90 & \phs0.05 &  $-$0.24 \\
May 16 & 1998.3704 & 48.54 & 176.96 & 0.84 & 0.27 & 165.4 & 48.64 & 176.80 & $-$0.10 &   \phs0.16 \\
Jul 04 & 1998.5046 & 48.51 & 188.19 & 0.95 & 0.27 & 171.9 & 48.45 & 188.11 & \phs0.06 &   \phs0.08 \\
Oct 07 & 1998.7646 & 41.40 & 212.85 & 1.04 & 0.34 & 118.7 & 41.52 & 213.13 & $-$0.12 &  $-$0.28 \\
Feb 28 & 1999.1589 & 15.79 & 330.10 & 1.08 & 0.28 &  15.7 & 15.98 & 330.84 & $-$0.19 &  $-$0.74 \\
Apr 30 & 2004.3280 & 44.16 & 156.39 & 0.92 & 0.44 & 169.8 & 43.84 & 155.99 & \phs0.32 &   \phs0.40 \\
May 01 & 2004.3307 & 43.77 & 156.12 & 0.84 & 0.46 & 160.8 & 43.93 & 156.27 & $-$0.16 &  $-$0.15 \\
May 03 & 2004.3362 & 44.11 & 156.74 & 0.87 & 0.45 & 160.2 & 44.13 & 156.84 & $-$0.02 &  $-$0.10 \\
May 04 & 2004.3389 & 44.19 & 157.19 & 0.84 & 0.47 & 155.0 & 44.22 & 157.11 & $-$0.03 &   \phs0.08 \\
May 05 & 2004.3417 & 44.28 & 157.45 & 0.83 & 0.46 & 159.4 & 44.31 & 157.40 & $-$0.03 &   \phs0.05 \\
May 06 & 2004.3444 & 44.31 & 157.65 & 0.83 & 0.46 & 159.4 & 44.41 & 157.67 & $-$0.10 &  $-$0.02 \\
Jul 07 & 2004.5142 & 48.31 & 173.28 & 0.49 & 0.19 &  99.3 & 48.26 & 173.32 & \phs0.05 &  $-$0.04 \\
Jul 08 & 2004.5168 & 48.36 & 173.64 & 0.42 & 0.22 & 104.0 & 48.29 & 173.55 & \phs0.07 &   \phs0.09 \\
Jul 09 & 2004.5197 & 48.39 & 173.82 & 0.33 & 0.32 & 127.3 & 48.33 & 173.79 & \phs0.06 &   \phs0.03 \\
Jul 21 & 2004.5525 & 49.56 & 175.39 & 1.01 & 0.45 & 139.7 & 48.63 & 176.58 & \phs0.93 &  $-$1.19 \\
Jul 23 & 2004.5580 & 48.72 & 177.01 & 0.98 & 0.41 & 152.4 & 48.67 & 177.05 & \phs0.05 &  $-$0.04 \\
Jul 30 & 2004.5771 & 48.87 & 178.74 & 0.96 & 0.43 & 149.5 & 48.77 & 178.66 & \phs0.10 &   \phs0.08 \\
Jul 31 & 2004.5800 & 49.22 & 178.79 & 1.06 & 0.42 & 132.5 & 48.78 & 178.90 & \phs0.44 &  $-$0.11 \\
May 24 & 2005.3929 & 15.17 &  11.44 & 0.49 & 0.25 &   7.0 & 15.45 &  12.34 & $-$0.28 &  $-$0.90 \\
Jun 09 & 2005.4368 & 17.16 &  46.12 & 0.52 & 0.25 & 174.0 & 16.99 &  46.31 & \phs0.17 &  $-$0.19 \\
Jun 16 & 2005.4561 & 18.07 &  59.39 & 0.53 & 0.24 &   2.6 & 18.11 &  58.88 & $-$0.04 &   \phs0.51 \\

\enddata
\tablecomments{Col. (1): UT month and day of NPOI observation 
Col. (2): Julian Year of NPOI observation. Col. (3): Fitted 
binary separation. Col. (4): Fitted binary position angle. Col. (5): Semimajor 
axis of error ellipse. Col. (6): Semiminor axis of error ellipse. 
Col. (7): Position angle of error ellipse.  
Col. (8): Calculated binary separation. Col. (9): Calculated binary position angle 
Col. (10): O$-$C value for binary separation. Col. (11): O$-$C value for binary 
position angle.}

\end{deluxetable}

\clearpage

\begin{deluxetable}{cc}
\tablecaption{O{\sc rbital} E{\sc lements and} S{\sc tellar} P{\sc arameters} \label{ephem}}
\tablewidth{0pt}
\tablehead{ \colhead{Data} & \colhead{Value}}
\startdata
a (mas)  & 32.1 $\pm$ 0.2 \\
i (deg)  & 12.1 $\pm$ 2.9 \\
$\Omega$ (deg) & 9.1 $\pm$ 2.5 \\
e        & 0.522 $\pm$ 0.004 \\
$\omega$ (deg)& 351.9 $\pm$ 2.7 \\
T$_0$ (JD)   & 2450121.8 $\pm$ 1.0 \\
P (days) & 564.69 $\pm$ 0.13 \\
M$_1 + $M$_2$ (M$_{\odot}$) & 4.7 $\pm$ 0.6 \\
$\gamma$ (km sec$^{-1}) $ & $-$16.66 $\pm$ 0.05 \\
K$_1$ (km sec$^{-1}) $ & 2.5 \\
K$_2$ (km sec$^{-1}) $ & 8.1 \\
$\chi_\nu^2$ & 0.8 \\
D$_{\rm A}$ (mas) & 0.4 \\
D$_{\rm B}$ (mas) & 0.2 \\
\enddata
\tablecomments{See \S 3.1 for a discussion of the methods used to obtain these results.
The stellar diameters (D$_{A,B}$) were fixed as discussed in that section.}  
\end{deluxetable}

\clearpage

\begin{deluxetable}{cccc}
\tablecaption{M{\sc agnitude} D{\sc ifference} R{\sc esults} \label{magdiff}}
\tablewidth{0pt}
\tablehead{ 
\colhead{Year} & \colhead{$\Delta$mag(5500\AA)} & \colhead{$\Delta$mag(7000\AA)} 
& \colhead{ Number of Epochs}}
\startdata
1997 & 2.54 $\pm$ 0.02 & 2.42 $\pm$ 0.01 & 5 \\
1998 & 2.58 $\pm$ 0.03 & 2.43 $\pm$ 0.01 & 3 \\
1999 & 2.57 $\pm$ 0.05 & 2.36 $\pm$ 0.02 & 1 \\
2004 & 2.56 $\pm$ 0.01 & 2.40 $\pm$ 0.01 & 14 \\
2005 & 2.58 $\pm$ 0.02 & 2.37 $\pm$ 0.01 & 3 \\
1997$-$2005 & 2.57 $\pm$ 0.05 & 2.39 $\pm$ 0.05 & 26 \\
\enddata
\tablecomments{The data for the individual years is plotted in 
Fig.~\ref{delmag}. See \S3.1 for an explanation of the net result 
shown for the years 1997$-$2005.}  
\end{deluxetable}

\clearpage

\begin{deluxetable}{cllllcccccr}
\tabletypesize{\scriptsize}
\rotate
\tablecaption{T{\sc he} R{\sc adial} V{\sc elocities of} \ph\ B \label{rv-b}}
\tablewidth{0pt}
\tablehead{
\colhead{ Central} & \colhead{UT Date} & \colhead{Epoch\tablenotemark{a}} & \colhead{Ident.} & \colhead {Laboratory} 
& \colhead{W$_{\lambda}$} & \colhead{Line Depth} & \colhead{Line Width} 
& & \colhead{RV (km sec$^{-1}$)} \\ 
\colhead{ ${\lambda}$ (\AA)} & & & \colhead{ Wavelength (\AA)} & \colhead{(\AA)} & \colhead{(m\AA)} & 
& \colhead{(\AA)} &  \colhead{Observed\tablenotemark{c}} & \colhead{Pred.} & \colhead{Diff.} }
\startdata
4864 & 2004 Jun 10 & 3166.7581\tablenotemark{b} & Ti II(82) & 4805.104 & 13 & 0.011 & 1.3 & -17.6 & -13.8 & -3.8 \\
     & & & Fe II(42) & 4923.930 & 23 & 0.020 & 1.3 & -11.8 & -13.8 & 2.0  \\
     & & & Ba II(1)  & 4934.086  & 11 & 0.009 & 1.3 & -17.3 & -13.8 & -3.5 \\
5140 & 2004 Jul 28 & 3214.7204 & Mg I(2)   & 5167.322  & 40 & 0.032 & 1.4 & -8.1  & -13.4 & 5.3 \\
     & & & Mg I(2)   & 5183.604  & 32 & 0.026 & 1.4 & -10.4 & -13.4 & 3.0 \\
5278 & 2005 Jun 24 & 3575.7405 & Fe II(J)  & 5279.03   & 36 & 0.026 & 1.5 & -26.0 & -19.7 & -6.3 \\
\enddata
\tablenotetext{a}{Epoch = HJD $-$ 2450000.0}
\tablenotetext{b}{Clouds caused a loss of 25 minutes of observing time.}
\tablenotetext{c}{Errors are $\pm$ 5 km sec$^{-1}$}
\end{deluxetable}

\clearpage

\begin{deluxetable}{lcccccl}
\tablecaption{M{\sc icroturbulence} D{\sc eterminations from}  Fe II L{\sc ines} 
\label{micro}}
\tablewidth{0pt}
\tablehead{
        & \colhead{Number} &   \colhead{$\xi_{1}$} &  & \colhead{$\xi_{2}$}   &  & \\
\colhead{Species} &\colhead{of Lines}&\colhead{(km s$^{-1}$)}&\colhead{log N/N$_{T}$}
&\colhead{(km s$^{-1}$)}&\colhead{log N/N$_{T}$}&\colhead{gf values} }
\startdata
Fe II  & 56     & 0.3 & -4.33$\pm$0.19 & 0.5 & -4.34$\pm$0.19 & MF+N4\\
        & 182    & 0.3 & -4.34$\pm$0.17 & 0.6 & -4.36$\pm$0.17 & MF+N4+KX\\
       & adopted& 0.4 &                &     &                &    \\
\enddata
\tablecomments{gf value references: MF = Fuhr et al. (1988), KX = Kurucz \& 
Bell(1995), N4 = Fuhr \& Wiese (2005). For $\xi_{1}$ and $\xi_{2}$ the abundances 
are found so that there is no trend of values for lines of different equivalent widths 
and have minimum scatter, respectively.}
\end{deluxetable}

\clearpage

\begin{deluxetable}{lcrrrr}
\tablecaption{A{\sc bundances of} $\phi$ H{\sc er} I{\sc ncluding} N{\sc ew} L{\sc ines} 
              \label{abun}}
\tablewidth{0pt}
\tablehead{
\colhead{Mult.} & \colhead{$\lambda$({\AA})}& \colhead{log gf} & \colhead{Ref.} & 
\colhead{W$_{\lambda}$(m{\AA})} & \colhead{log N/N$_{T}$} }
\startdata
C II& & & \multicolumn{3}{c}{log C/N$_{T}$ = -3.71$\pm$0.20}\\
O I& & & \multicolumn{3}{c}{log O/N$_{T}$ = -3.37$\pm$0.07}\\
12 & 5329.11 & -1.24 & WF &   8 & -3.38\\
    & 5329.68 & -1.02 & WF & 13 & -3.40\\
     & 5330.74 & -0.87 & WF  &18 & -3.26\\ 
Mg I& & & \multicolumn{3}{c}{log Mg/N$_{T}$ = -4.49$\pm$0.15}\\
   2 & 5167.32 & -1.03 & WS &  11 & -4.54\\
     & 5172.68 & -0.38 & WS &  34 & -4.53\\
     & 5183.60 & -0.16 & WS &  35 & -4.71\\
Mg II& & & \multicolumn{3}{c}{log Mg/N$_{T}$ = -4.76$\pm$0.03}\\
Al II& & & \multicolumn{3}{c}{log Al/N$_{T}$ = -6.28}\\
Si II& & & \multicolumn{3}{c}{log Si/N$_{T}$ = -4.83$\pm$0.17}\\
S II& & & \multicolumn{3}{c}{log S/N$_{T}$ = -4.72$\pm$0.21}\\
   1 & 4991.97 & -0.25 & KX &  8 & -4.54\\
   7 & 4925.34 & -0.47 & WS &  5 & -4.59\\
     & 5009.56 & -0.09 & WM &  8 & -4.65\\
     & 5032.47 & +0.18 & WS &  9 & -4.83\\
   9 & 4815.55 & +0.18 & WM &  8 & -5.02\\
  15 & 5014.04 & +0.03 & WM &  8 & -4.56\\
  38 & 5320.73  & +0.46 & WS &  6 & -4.66\\
& 5345.72 & +0.28 & WS  &  6 & -4.43\\
  39 & 5212.62 & +0.24 & WS &  5 & -4.57\\ 
Ca I& & & \multicolumn{3}{c}{log Ca/N$_{T}$ = -5.14}\\
Ca II& & & \multicolumn{3}{c}{log Ca/N$_{T}$ = -5.51$\pm$0.11}\\
    2 & 8498.02 & -1.31 & WS & 109 & -5.37\\
      & 8542.09 & -0.36 & WS & 179 & -5.46\\ 
& 8662.14 & -0.62 & WS & 136 & -5.67\\
Sc II& & & \multicolumn{3}{c}{log Sc/N$_{T}$ = -7.33$\pm$0.14}\\
  23 & 5031.02 & -0.32 & LD & 33 & -7.37\\
Ti II& & & \multicolumn{3}{c}{log Ti/N$_{T}$ = -6.21$\pm$0.21}\\
    7 & 5154.07 & -1.92 & MF &  15 & -6.10\\
       & 5188.68 & -1.21 & MF &  43 & -5.98\\
  13 & 5010.21 & -1.34 & KX &   9 & -6.14\\
  17 & 4798.52 & -2.43 & MF &   5 & -6.44\\
  69 & 5336.78 & -1.70 & MF & 20 & -6.15\\
 70 & 5226.54 & -1.30 & MF & 41 & -5.95\\
   & 5262.13 & -2.11 & KX & 11 & -6.10\\
 71 & 5013.68 & -1.94 & KX &   8 & -6.40\\
  82 & 4805.09 & -1.10 & MF &  31 & -6.17\\
  86 & 5129.15 & -1.39 & MF &  25 & -6.15\\
     & 5185.91 & -1.35 & MF &  24 & -6.21\\
103 & 5211.54 & -1.36 & KX & 10 & -6.35\\
       & 5268.62 & -1.62 & MF &   7 & -6.28\\ 
 113 & 5072.28 & -0.75 & MF &  16 & -6.42\\
 114 & 4874.01 & -0.79 & MF &  15 & -6.44\\
V II& & & \multicolumn{3}{c}{log V/N$_{T}$ = -8.07}\\
Cr I& & & \multicolumn{3}{c}{log Cr/N$_{T}$ = -5.12$\pm$0.18}\\
   7 & 5204.51 & -0.21 & MF &   8 & -5.04\\
     & 5206.03 & +0.02 & MF &  11 & -5.14\\
     & 5208.42 & +0.16 & MF &  16 & -5.05\\
Cr II& & & \multicolumn{3}{c}{log Cr/N$_{T}$ = -5.25$\pm$0.19}\\
 23 & 5246.75 & -2.45 & MF &  24 & -5.15\\
    & 5249.40 & -2.43 & KX &  21 & -5.24\\ 
    & 5318.41 & -3.13 & KX &    7 & -5.15\\
 24 & 5210.87 & -2.94 & KX &  14 & -5.00\\
      & 523.250 & -2.09 & KX &  27 & -5.20\\
      & 5305.84 & -2.36 & KX &  36 & -4.82\\
   30 & 4812.34 & -1.80 & MF &  33 & -5.48\\
        & 4836.22 & -2.25 & MF &  36 & -4.93\\
        & 4848.24 & -1.14 & MF &  62 & -5.22\\
        & 4856.19 & -2.26 & MF &  16 & -5.54\\
        & 4884.60 & -2.08 & MF &  30 & -5.31\\
  43 & 5232.50 & -2.09 & KX &  28 &  -5.20\\
       & 5237.33 & -1.16 & MF &  62 & -5.07\\ 
       & 5274.96 & -1.29 & KX &  56 & -5.15\\  
       & 5279.88 & -2.10 & MF &  36 & -4.94\\
       & 5280.05 & -2.01 & KX  & 19 & -5.57\\
       & 5308.41 & -1.81 & MF &  35 & -5.28\\
       & 5310.69 & -2.28 & MF &  23 & -5.15\\ 
       & 5313.56 & -1.65 & MF &  46 & -5.09\\ 
       & 5334.87 & -1.56 & KX &  45 & -5.22\\
 190 & 4901.62 & -0.83 & KX &  21 & -5.45\\
     & 4912.46 & -0.95 & KX &  18 & -5.45\\
Mn I& & & \multicolumn{3}{c}{log Mn/N$_{T}$ = -4.91$\pm$0.19}\\
  16 & 4823.52 & +0.14 & MF &  13 & -4.88\\
Mn II& & & \multicolumn{3}{c}{log Mn/N$_{T}$ = -5.01$\pm$0.21}\\
   I & 4806.82 & -1.56 & KX &  31 & -5.04\\
     & 4811.62 & -2.34 & KX &   6 & -5.25\\
     & 4830.06 & -1.85 & KX &  13 & -4.98\\
     & 4839.74 & -1.86 & KX &  11 & -5.09\\
     & 4842.33 & -2.01 & KX &  12 & -4.85\\
     & 4847.60 & -1.81 & KX &  17 & -5.23\\
     & 5177.65 & -1.77 & KX &  23 & -4.75\\
     & 5251.82 & -1.83 & KX &     9 & -4.76\\
  & 5295.40 & -0.66 & KX &  42 & -4.69\\
   & 5302.44 & -1.00 & KX &  58 & -4.57\\
     & 6609.26 & -2.05 & KX &   4 & -4.94\\
     & 6682.38 & -3.11 & KX &   6 & -5.12\\
Fe I& & & \multicolumn{3}{c}{log Fe/N$_{T}$ = -4.14 $\pm$0.20}\\
   15 & 5328.05 &  -1.47 & N4 & 10 & -3.82\\ 
  289 & 4871.31 & -0.36 & N4 &   9 & -3.94\\
     & 4872.13 & -0.57 & N4 &   6 & -3.91\\
     & 4891.49 & -0.11 & N4 &   8 & -4.28\\
     & 4957.30 & -0.41 & N4 &   9 & -3.89\\
     & 4957.60 &  0.23 & N4 &  21 & -4.07\\
 383 & 5232.95 & -0.06 & N4 & 20 & -3.75\\ 
 553 & 5324.18 & -0.10 & N4 &  5 & -4.33\\ 
 638 & 5014.94 & -0.30 & N4 &  5 & -3.72\\
 984 & 5005.71 & -0.18 & KX &  7 & -3.78\\
Fe II& & & \multicolumn{3}{c}{log Fe/N$_{T}$ = -4.34$\pm$0.18}\\
  31 & 4893.83 & -4.27 & N4 &   3 & -4.52\\
 198 & 6416.92 & -2.88 & N4 &  20 & -4.25\\
   J & 4826.68 & -0.44 & KX &   5 & -4.47\\
     & 4883.28 & -0.64 & KX &   5 & -4.31\\
     & 4908.15 & -0.30 & KX &   8 & -4.36\\
     & 4913.29 & +0.01 & KX &  13 & -4.40\\
     & 4948.10 & -0.32 & KX &   6 & -4.46\\
     & 4948.79 & -0.01 & KX &  10 & -4.52\\
     & 4951.58 & +0.18 & KX &  15 & -4.45\\
     & 4953.98 & -2.76 & KX &   6 & -4.25\\
     & 4958.82 & -0.65 & KX &   3 & -4.43\\
     & 4977.03 & +0.04 & KX &  12 & -4.42\\
     & 4984.49 & +0.01 & KX &  16 & -4.23\\
     & 4990.50 & +0.18 & KX &  16 & -4.41\\
     & 4991.44 & -0.57 & KX &   7 & -4.17\\
     & 4993.35 & -3.65 & MF &  16 & -4.23\\
     & 5001.95 & +0.90 & KX &  44 & -4.18\\
     & 5004.19 & +0.50 & KX &  24 & -4.43\\
     & 5006.84 & -0.43 & KX &  10 & -4.09\\
     & 5007.45 & -0.37 & KX &   9 & -4.22\\
     & 5007.74 & -0.20 & KX &  10 & -4.34\\
     & 5009.02 & -0.42 & KX &   6 & -4.40\\
     & 5015.76 & -0.05 & KX &  22 & -3.94\\
     & 5018.44 & -1.22 & MF &  95 & -4.25\\
     & 5019.46 & -2.70 & KX &  12 & -3.97\\
     & 5021.59 & -0.30 & KX &  10 & -4.24\\
     & 5022.79 & -0.02 & KX &  14 & -4.33\\
     & 5026.80 & -0.22 & KX &  10 & -4.33\\
     & 5030.63 & +0.40 & KX &  21 & -4.43\\
     & 5031.90 & -0.78 & KX &   3 & -4.33\\
     & 5032.71 & +0.11 & KX &  14 & -4.41\\
     & 5035.70 & +0.61 & KX &  36 & -4.13\\
     & 5045.11 & -0.13 & KX &   9 & -4.48\\
     & 5060.26 & -0.52 & KX &   8 & -4.07\\
     & 5061.72 & +0.22 & KX &  16 & -4.44\\
     & 5067.89 & -0.20 & KX &   5 & -4.66\\
     & 5070.90 & +0.24 & KX &  16 & -4.45\\
     & 5075.76 & +0.28 & KX &  11 & -4.68\\
     & 5076.61 & -0.71 & KX &   7 & -4.00\\
     & 5082.23 & -0.10 & KX &   6 & -4.64\\
     & 5087.30 & -0.49 & KX &   3 & -4.58\\
     & 5093.57 & +0.11 & KX &  18 & -4.23\\
     & 5106.11 & -0.28 & KX &   9 & -4.29\\
     & 5117.03 & -0.13 & KX &   8 & -4.46\\
     & 5119.34 & -0.56 & KX &   4 & -4.38\\
     & 5120.35 & -4.21 & KX &   5 & -4.30\\
     & 5127.86 & -2.54 & KX &  10 & -4.23\\
     & 5144.35 & +0.28 & KX &  12 & -4.60\\
     & 5145.77 & -0.40 & KX &  12 & -3.96\\
     & 5149.46 & +0.40 & KX &  18 & -4.45\\
     & 5150.49 & -0.12 & KX &   7 & -4.51\\
     & 5160.84 & -2.64 & KX &  10 & -4.13\\
     & 5166.55 & -0.03 & KX &  11 & -4.35\\
     & 5169.03 & -0.87 & MF & 104 & -4.27\\
     & 5170.77 & -0.36 & KX &   8 & -4.22\\
     & 5177.02 & -0.18 & KX &  10 & -4.29\\
     & 5180.31 & +0.04 & KX &  11 & -4.46\\
     & 5186.87 & -0.30 & KX &   4 & -4.60\\
     & 5194.89 & -0.15 & KX &  13 & -4.10\\
     & 5199.12 & +0.10 & KX & 13 & -4.44\\
     & 5200.80 & -0.37 & KX &   6 & -4.39\\
     & 5203.64 & -0.05 & KX &   8 & -4.54\\
     & 5214.00 & -0.22 & KX &  15 & -3.95\\
     & 5214.05 & -0.90 & KX &    5 & -3.90\\
     & 5215.34 & -0.10 & KX &  16 & -4.06\\  
     & 5215.83 & -0.23 & KX &   17 & -3.92\\
     & 5216.85 & +0.81& KX &   30 & -4.39\\   
     & 5218.84 & -0.20 & KX &     7 & -4.48\\
     & 5222.36 & -0.33 & KX &     6 & -4.39\\
     & 5223.23 & -0.41 & KX &   10 & -4.07\\
     & 5223.80 & -0.59 & KX &     6 & -4.20\\
     & 5224.40 & -0.57 & KX &     5 & -4.23\\ 
     & 5227.48 & +0.85 & N4 &  44 & -4.02\\ 
     & 5231.91 & -0.64 & KX &     5 & -4.15\\
     & 5234.62 & -2.05 & MF &   62 & -4.14\\
     & 5237.95 & +0.14 & KX &   16 & -4.28\\
     & 5245.46 & -0.51 & KX &     5 & -4.34\\
     & 5247.95 & +0.55 & N4&  21 & -4.44\\
     & 5251.23 &  0.42 & N4 &   24 & -4.19\\
     & 5253.64 & -0.09 & KX &  12 & -4.22\\
     & 5254.41 & -0.77 & KX &    6 & -3.96\\
     & 5254.93 & -3.23 & KX  &  21 & -4.25\\
     & 5257.12 & +0.03 & KX &  13 & -4.25\\
     & 5260.26 & +1.07 & KX &  40 & -4.37\\
     & 5264.18 & +0.30 & N4 &  24 & -4.08\\
     & 5264.81 & -3.19 & MF &  29 & -4.04\\
     & 5272.40 & -2.03 & MF &  19 & -4.14\\
     & 5276.00 & -1.94 & MF &   6  & -4.30\\
     & 5278.21 & -1.56 & KX &   5 & -4.24\\
     & 5278.94 & -2.41 & KX &   8 & -4.32\\
     & 5291.67 & +0.58 & KX &  21 & -4.49\\
     & 5303.39 & -1.61 & KX  &    4 & -4.30\\
     & 5306.18 & +0.04 & N4 &  12 & -4.31\\
     & 5315.08 & -0.38 & KX  &    5 & -4.34\\
     & 5315.56 & -1.46 & KX  &   6 & -4.28\\
     & 5316.23 & +0.34 & N4 & 26 & -4.10\\
     & 5316.78 & -2.78 & N4 &  25 & -4.58\\
     & 5318.05 & -0.14 & KX &   7 & -4.46\\
     & 5318.75 & -0.57 & KX &   4 & -4.34\\
     & 5339.59 &  0.54 & KX &  23 & -4.36\\
     & 5347.18 & -0.28 & KX &    6 & -4.33\\
Fe III& & & \multicolumn{3}{c}{log Fe/N$_{T}$ = -4.32}\\
Ni II& & & \multicolumn{3}{c}{log Ni/N$_{T}$ = -6.14$\pm$0.14}\\
Zn I& & & \multicolumn{3}{c}{log Zn/N$_{T}$ = -5.66$\pm$0.18}\\
   2 & 4810.53 & -0.14 & KX &  18 & -5.53\\
Zn II& & & \multicolumn{3}{c}{log Zn/N$_{T}$ = -5.56}\\
   3 & 4911.18 & +0.54 & WM &  18 & -5.56\\
Ga II& & & \multicolumn{3}{c}{log Ga/N$_{T}$ = -5.87$\pm$0.10}\\
Sr II& & & \multicolumn{3}{c}{log Sr/N$_{T}$ = -7.99$\pm$0.08}\\
Y II& & & \multicolumn{3}{c}{log Y/N$_{T}$ = -6.65$\pm$0.17}\\
  20 & 4982.13 & -1.29 & HL &  20 & -6.58\\
     & 5119.11 & -1.36 & HL &  18 & -6.62\\
     & 5200.40 & -0.57 & HL &  40 & -6.60\\
     & 5205.72 & -0.34 & HL &  52 & -6.27\\
     & 5289.82 & -1.85 & HL &    9 &  -6.55\\
  22 & 4823.30 & -1.11 & HL &  29 & -6.49\\
  22 & 4854.86 & -0.38 & HL &  36 & -6.97\\
     & 4883.68 &  0.07 & HL &  58 & -6.39\\
  28 & 5196.42 & -0.88 & KX &  17 & -6.73\\
  38 & 6613.73 & -1.11 & HL &  12 & -6.70\\
Zr II& & & \multicolumn{3}{c}{log Zr/N$_{T}$ = -7.24$\pm$0.20}\\
Ba II& & & \multicolumn{3}{c}{log Ba/N$_{T}$ = -8.36}\\
   1 & 4934.08 &  0.00 & WM &  17 & -8.36\\
Ce II& & & \multicolumn{3}{c}{log Ce/N$_{T}$ = -7.61}\\
Hg I& & & \multicolumn{3}{c}{log Hg/N$_{T}$ = -6.14}\\
Hg II& & & \multicolumn{3}{c}{log Hg/N$_{T}$ = -6.19}\\
\enddata
\tablecomments{gf value references (including those for older lines): 
BG = \citet{bg89} for V II; \citet{bg81} for Zr II,
DW = Dworetsky (1980), GB = Grevesse et al. (1981), HL = Hannaford et al.
(1982),
KX = Kurucz \& Bell (1995), LA = Lanz \& Artru (1985), LD = Lawler \& Dakin
(1989),
MF = Fuhr et al. (1988) and Martin et al. (1988), MC = Magazzu \& Cowley
(1986),
N4 = Fuhr \& Wiese (2005), WF = Wiese, Fuhr \& Deters (1996),WM = Wiese \& Martin
(1980),
WS = Wiese, Smith \& Glennon (1966) and Wiese, Smith \& Miles (1969)
The adopted abundance from Hg II $\lambda$3984 is from Woolf \& Lambert
(1999b). 
The multiplet numbers are from Moore (1945) except that I indicates  Mn II lines 
from Iglesias \& Velasco (1964)  and J indicates Fe II lines from Johansson (1978 ).
 }
\end{deluxetable}

\clearpage

\begin{deluxetable}{llllr}
\tabletypesize{\scriptsize}
\tablecaption{C{\sc omparison of} D{\sc erived and} S{\sc olar} 
              A{\sc bundances} (log N/H) \label{compare}}
\tablewidth{0pt}
\tablehead{ 
   & \colhead{$\phi$ Her A} & \colhead{$\phi$ Her A} & \colhead{Number}& \\
\colhead{Species} & \colhead{Adelman et al. (2001)} & \colhead{This Paper} & 
\colhead{of Lines}   & \colhead{Sun} }
\startdata
He I   & -1.21$\pm$0.05 & -1.17$\pm$0.05 &  7 & -1.01\\
C II   & -3.76$\pm$0.25 & -3.68$\pm$0.20 &  3 & -3.45\\
O I    & -3.17$\pm$0.49 & -3.34$\pm$0.07 &  4 & -3.13\\
Mg I   & -4.54$\pm$0.41 & -4.46$\pm$0.15 &  5 & -4.42\\
Mg II  & -4.78$\pm$0.03 & -4.75$\pm$0.03 &  5 & -4.42\\
Al II  & -6.27          & -6.25          &  1 & -5.53\\
Si II  & -4.95$\pm$0.22 & -4.80$\pm$0.17 & 11 & -4.45\\
S II   & -4.41$\pm$0.25 & -4.69$\pm$0.21 & 22 & -4.67\\
Ca I   & -5.14          & -5.11          &  1 & -5.64\\
Ca II  & -5.56          & -5.48$\pm$0.11 &  4 & -5.64\\
Sc II  & -7.39$\pm$0.14 & -7.03$\pm$0.14 & 12 & -8.83\\
Ti II  & -6.32$\pm$0.29 & -6.21$\pm$0.21 & 66 & -6.98\\
V II   & -8.11          & -8.04          &  1 & -8.00\\
Cr I   & -5.16$\pm$0.28 & -5.09$\pm$0.18 &  7 & -6.33\\
Cr II  & -5.39$\pm$0.25 & -5.22$\pm$0.19 & 64 & -6.33\\
Mn I   & -4.86$\pm$0.22 & -4.88$\pm$0.19 & 18 & -6.61\\
Mn II  & -4.95$\pm$0.29 & -4.98$\pm$0.21 & 71 & -6.61\\
Fe I   & -4.19$\pm$0.22 & -4.11$\pm$0.20 & 70 & -4.50\\
Fe II  & -4.45$\pm$0.20 & -4.31$\pm$0.18 &182 & -4.50\\
Fe III & -4.35          & -4.29          &  1 & -4.50\\
Ni II  & -6.30$\pm$0.30 & -6.11$\pm$0.14 &  4 & -5.75\\
Zn I   & -5.80          & -5.63$\pm$0.18 &  2 & -7.40\\
Zn II  & -5.61          & -5.53          &  1 & -7.40\\
Ga II  & -5.85$\pm$0.18 & -5.84$\pm$0.10 &  2 & -9.12\\
Sr II  & -8.05$\pm$0.14 & -7.96$\pm$0.08 &  3 & -9.03\\
Y II   & -6.79$\pm$0.23 & -6.62$\pm$0.17 & 21 & -9.76\\
Zr II  & -7.30$\pm$0.25 & -7.21$\pm$0.20 & 33 & -9.40\\
Ba II  & -7.69          & -8.33          &  1 & -9.87\\
Ce II  & -7.63          & -7.58          &  1 & -10.42\\
Pr     &     ....       &-9.60           & ... & -11.29\\
Nd     &     ....       & -9.30          & ... & -10.59\\
Hg I   & -6.15          & -6.11          &  1 &-10.83\\
Hg II  & -6.41          & -6.16          &  1 &-10.83\\
\enddata
\tablecomments{The Hg II value adopted for this study is that of
Woolf \& Lambert (1999b), the Nd and Pr abundances are from
Dolk et al. (2002), and the solar values are from Grevesse,
Noels \& Sauval (1996). When more than one line is detected 
the error is the standard deviation of the mean. No error is 
quoted when only one line is present.}
\end{deluxetable}
 
\clearpage

\begin{deluxetable}{cccc}
\tablecaption{S{\sc tellar} P{\sc arameters of} \ph\ A {\sc and} B 
\label{stars}}
\tablewidth{0pt}
\tablehead{\colhead{Component} & \colhead{T$_{eff}$} & \colhead{log (g)} & \colhead{vsin(i)} \\
            & \colhead{(K)} & \colhead{log(cm sec$^{-2}$)} & \colhead{(km sec$^{-1}$)} }
\startdata
A & 11525 $\pm$ 150 & 4.05 $\pm$ 0.15 & 8.0 $\pm$ 1.0 \\
B &  8000 $\pm$ 150 & 4.30 $\pm$ 0.15 & 50.0 $\pm$ 3.0 \\
\enddata
\end{deluxetable}

\end{document}